\documentclass[aps,prd,twocolumn,showpacs,groupedaddress,nofootinbib,letterpaper,floatfix]{revtex4-1}

\usepackage{dcolumn}
\usepackage{amsmath}
\usepackage{amssymb}
\usepackage{dsfont}
\usepackage{graphicx}
\usepackage{hyperref}
\usepackage{braket}
\usepackage{color}
\usepackage{bbm}
\usepackage{mathtools}

\DeclareMathOperator{\sinc}{sinc}
\DeclareMathOperator{\Ci}{Ci}
\DeclareMathOperator{\Si}{Si}

\DeclareMathOperator{\tr}{tr}
\DeclareMathOperator{\spec}{spec}

\DeclarePairedDelimiter{\avg}{\langle}{\rangle}
\DeclarePairedDelimiter{\abs}{\lvert}{\rvert}

\begin{document}

\title{Locality and entanglement in bandlimited quantum field theory}

\author{Jason Pye}
 \email{j2pye@uwaterloo.ca}
\affiliation{Department of Applied Mathematics, University of Waterloo, Waterloo, Ontario, N2L 3G1, Canada}

\author{William Donnelly}
 \email{donnelly@physics.ucsb.edu}
\affiliation{Department of Physics, University of California Santa Barbara, Santa Barbara, California 93106, USA}

\author{Achim Kempf}
\email{akempf@uwaterloo.ca}
\affiliation{Departments of Applied Mathematics and Physics and Institute for Quantum Computing, 
University of Waterloo, Waterloo, Ontario, N2L 3G1, Canada }

\date{\today}

\begin{abstract}
We consider a model for a Planck scale ultraviolet cutoff which is based on Shannon sampling. Shannon sampling originated in information theory, where it expresses the equivalence of continuous and discrete representations of information. When applied to quantum field theory, Shannon sampling expresses a hard ultraviolet cutoff in the form of a bandlimitation. This introduces nonlocality at the cutoff scale in a way that is more subtle than a simple discretization of space: quantum fields can then be represented as either living on continuous space or, entirely equivalently, as living on any one lattice whose average spacing is sufficiently small.   
We explicitly calculate vacuum entanglement entropies in 1+1 dimension and we find a transition between logarithmic and linear scaling of the entropy, which is the expected 1+1 dimensional analog of the transition from an area to a volume law.  
We also use entanglement entropy and mutual information as measures to probe in detail the localizability of the field degrees of freedom. We find that, even though neither translation nor rotation invariance are broken, each field degree of freedom occupies an incompressible volume of space, indicating a finite information density.
\end{abstract}
\pacs{
03.67.-a, 
03.70.+k, 
04.60.-m, 
89.70.-a  
}
\maketitle


\section{Introduction}
\label{section:intro}

It is generally thought that quantum theory and general relativity, when combined, imply the existence of a minimum length in nature, see e.g., \cite{Wheeler1957,Mead1964,Kempf1994,Garay1994, Witten1996,Hossenfelder:2012jw}.
One line of argument is to consider that when attempting to resolve distances with a smaller and smaller uncertainty $\Delta x$, the thereby necessarily increasing momentum uncertainty will induce a correspondingly increasing curvature uncertainty.
Eventually the growing curvature uncertainty should make it impossible to resolve distances any further, leading to $\Delta x \gtrsim \ell_P$, where $\ell_P$ is at the Planck scale, or perhaps the string scale.

Another line of argument comes from black hole physics. 
Black holes (and more general causal horizons \cite{Jacobson2003}) are believed to carry a finite entropy given by the Bekenstein-Hawking formula, $S_\text{BH} = \frac{A}{4 G}$ (we set $c = \hbar = 1$).
This entropy may have its origin in the entanglement entropy of quantum fields \cite{Sorkin1983,Bombelli1986,Srednicki1993} (see \cite{Solodukhin2011} for a review): the entanglement entropy has the same scaling behavior as $S_\text{BH}$ and the dependence on the number of particle species could also be matched when taking into account how the renormalized gravitational coupling depends on the number of species \cite{Jacobson:1994iw,Susskind1994}. Crucially, the entanglement entropy also matches the order of magnitude of $S_\text{BH}$, if there exists a natural ultraviolet cutoff which is close to the Planck length, $\ell_P = \sqrt{G}$. 

Given these and other arguments, (see e.g., \cite{Bianchi2012}), for the existence of a natural ultraviolet cutoff at the Planck length, the question arises as to the nature of this ultraviolet cutoff.
In which sense might the density of degrees of freedom in quantum fields be finite?
How are these degrees of freedom distributed and how can they be spatially resolved?  

To completely answer these questions would require knowledge of the still-to-be-developed theory of quantum gravity.
More feasible at the present stage is to study how the natural ultraviolet cutoff may first manifest itself as one approaches the Planck scale from low energies, i.e., when coming from the safe ground of low energy physics where conventional quantum field theory (QFT) still holds.
The question then is how can one model the first modifications to quantum field theory that express the impact of a Planck length cutoff?

One simple model is that of QFT on discretized space or spacetime. 
A difficulty with this approach is the associated complete breaking of local Poincar\'e symmetry, though there are interesting methods to generate discretizations whose statistics are Poincar\'e invariant \cite{Bombelli:1987aa,Bombelli2006,Henson2006}.
Also, when lattices evolve in time, there tend to be problems with non-adiabaticity and an associated excessive particle production from the vacuum, in particular, in an expanding universe \cite{Foster2004}.

Here, we will consider a different model for a natural ultraviolet cutoff in QFT with which at least the local Euclidean symmetries are preserved and which does not necessarily induce non-adiabaticities in an expanding universe.
The model is that of QFT with a hard cutoff in the form of a finite spatial bandwidth, i.e, a finite smallest wavelength \cite{Kempf:2000fc,Kempf:2000ac,Kempf:2006wp,Bojowald2012}.
How does this compare to the models of QFT on a lattice?
It is well-known that the discretization of space gives rise to a minimum wavelength.
It is less well-known that, vice versa, to impose a lower bound on wavelengths does not automatically lead to a discretization of space. 
Instead, one can also obtain a so-called sampling theoretic cutoff which does not break symmetries such as translation invariance. With this cutoff, the theory does not live on a single dedicated lattice.
Instead, it lives on continuous space and, entirely equivalently, it can be written as living on any one lattice whose average spacing is sufficiently small.

The mathematical structure that underlies the sampling theoretic ultraviolet cutoff is Shannon sampling theory.
Shannon sampling plays a central r\^ole in information theory because it establishes the equivalence of continuous and discrete representations of information \cite{Shannon1948}.

Consider, for example, a music signal, $f(t)$, that is low-pass filtered to some finite bandwidth, $\Omega$.
Even though the bandlimited music signal is a continuous function in time, it suffices to know the signal's amplitudes $f(t_n)$ on any single lattice of points $\{t_n\}$, if the lattice has an average spacing at or below the critical spacing $\pi/\Omega$, which is called the Nyquist spacing. 
From the amplitude samples $\{f(t_n)\}$, the signal $f(t)$ can then be reconstructed for all $t$, in principle, without error.
This establishes the equivalence of continuous and discrete representations of information. 
The fact that none of the set of sufficiently densely-spaced lattices is preferred allows the preservation of full translation invariance (and in higher dimensions also rotation invariance).

Applied to physics, this means that spacetime could be simultaneously continuous \it and \rm discrete in mathematically the same way that information (such as a music signal) possesses simultaneously both a continuous representation and equivalent discrete representations.
Physical fields then possess a representation on continuous space, while being fully equivalently represented also by their amplitudes on any one lattice of sufficiently dense spacing. The democracy among these lattices allows translation and rotation invariance and more generally also Killing vector fields to be preserved with the cutoff. 

Our aim here is to study the implications of this sampling theoretic ultraviolet cutoff for localization and entanglement in quantum field theory.

We find that the sampling-theoretic ultraviolet cutoff implies a particular kind of nonlocality in QFT which manifests itself through small but non-vanishing equal-time commutators at spacelike separation.
This type of nonlocality appears naturally in perturbative quantum gravity \cite{Giddings2015,Donnelly:2015hta} and in string field theory \cite{Lowe1995}.
We also show how correlation functions are affected by the bandlimit, both classically at finite temperature, and quantum-mechanically.

Further, in order to probe the spatial localization of degrees of freedom with this UV cutoff, we examine the behaviour of the entanglement entropy in a quantum field.
For comparison, recall that in a QFT regulated on a lattice, the local field oscillators are coupled via the discretized spatial Laplace operator in the Lagrangian. This makes their joint ground state, i.e., the vacuum state, an entangled state. Tracing over a region therefore yields an entanglement entropy.
Due to the generally short range of the vacuum entanglement, the contributions to the entanglement entropy are dominated by correlations between those local field oscillators that are close the boundary of the considered region. This usually leads to an area law for the entanglement entropy, see e.g., \cite{Eisert2008}.
In the 1+1 dimensional massless field theory that we will consider, the correlation length is infinite and the entanglement entropy therefore grows logarithmically (see \cite{Holzhey1994}).

Unlike in such lattice theories, a bandlimited QFT does not have a preferred lattice representation. This makes the splitting of space into distinct regions nontrivial and, as a consequence, the Hilbert space does not automatically factorize into subsystems. 
Instead, more subtly, any discrete subset of points in space now defines a subsystem, and one can calculate its entanglement entropy with the rest of the system.
In particular, we show how to calculate the entanglement entropy for a subset of degrees of freedom in a Gaussian state.
Our result generalizes known formulae for the entropy of Gaussian states to the setting where the commutation relations are nontrivial.

With samples placed at the Nyquist spacing, $\pi/\Omega$, we recover the usual logarithmic scaling behaviour of the entanglement entropy.
When the spacing is larger than the Nyquist spacing (undersampling), we find that the entanglement entropy crosses over to a volume law. Further, unlike in a lattice theory, in the bandlimited theory it is possible to probe the entanglement between degrees of freedom that are arbitrarily closely spaced.
Surprisingly, when the spacing of samples is smaller than the Nyquist spacing (oversampling), we find no reduction in entanglement entropy.
Since the sampling points can occupy an arbitrarily small region, this, naively, appears to indicate that the field still carries an infinite density of local degrees of freedom (local field oscillators).
We show that the resolution of this apparent paradox is that the region of space being probed by the sample points does not actually decrease as the samples are taken closer together.
In effect, each local field degree of freedom occupies an incompressible volume.

Finally, we briefly examine the infrared behaviour of the entanglement entropy in the 1+1 dimensional bandlimited theory.
In addition to the well-known logarithmic growth of the entanglement entropy, there is a subleading double logarithmic infrared divergence.
This divergence appears whenever there is a continuum of modes above the infrared cutoff scale; thus it appears when regulating the infrared with a mass or a hard momentum cutoff, but not if regulating by imposing periodic boundary conditions.

We conclude with a discussion of implications and directions for future work.

\section{Sampling theory for quantum fields}
\label{section:sampling}

\subsection{Overview of classical sampling theory}

The central result of classical sampling theory is Shannon's sampling theorem \cite{Shannon1948}.
This theorem establishes an equivalence between discrete and continuous representations of information and it is, therefore, in ubiquitous use in communication engineering and signal processing.
Let us consider a \emph{bandlimited} signal, $\phi(x)$, i.e., a signal whose Fourier transform is supported on an interval $(-\Omega, \Omega)$, whose width is the \emph{bandwidth} $2 \Omega$:
\begin{equation}
\phi(x) = \int_{-\Omega}^{\Omega} \frac{dk}{2\pi} \tilde{\phi}(k) e^{ikx}
\end{equation}
Shannon's theorem states that such a function is completely determined by its values $\phi(x_n^{(\alpha)})$ on a discrete lattice of points $\{x_n^{(\alpha)}\}$, where
$x_n^{(\alpha)} = (2 \pi n - \alpha)/(2\Omega)$ and $n\in \mathds{Z}$. Here, $\alpha \in [0,2\pi)$ is an arbitrary fixed constant that labels lattices.
Given the values, $\{\phi(x_n^{(\alpha)})\}$, of the function on this lattice, $\phi(x)$ can be recovered for any $x$ by the reconstruction formula:
\begin{equation}
\label{eq:shannon}
\phi(x) = \sum_{n \in \mathds{Z}} \sinc [ (x-x_n^{(\alpha)}) \Omega ]~ \phi(x_n^{(\alpha)})
\end{equation}
Thus, the space of bandlimited functions is completely equivalent to the space of functions defined on this lattice.
This has the practical implication of allowing us to work with concrete and therefore computationally-convenient representations of functions on a lattice (as well as quantum fields, as we will see below) while preserving translation invariance (because the function retains its equivalent continuous representation).

Another important finding in sampling theory is that the samples of the function do not have to be taken equidistantly: a bandlimited function can be reconstructed on any discrete set of points as long as these points are chosen with a sufficiently dense average spacing (technically, the Beurling average spacing \cite{Landau1967,Beurling1989}).
Here, this maximum average spacing is $\pi/\Omega$.
In general, such a set of points is called a \emph{sampling lattice}, and the case of equidistant samples with separation $\pi/\Omega$ is called a \emph{Nyquist lattice}.
In the case of sampling on a lattice other than a Nyquist lattice, perfect reconstruction is still possible but the reconstruction formula is more complicated than \eqref{eq:shannon} and the reconstruction becomes more sensitive to noise in the samples.
Because the degrees of freedom of the function (values at the sample points) are not confined to any particular sampling lattice, the information contained in bandlimited signal is subtly nonlocal.

Sampling theory generalizes readily to bandlimited functions in higher dimensions, whose Fourier transforms are supported in a compact region of momentum space, see \cite{Zayed1993,Higgins1996,Higgins1999,Benedetto2001} and it generalizes also to curved space \cite{Kempf2008}.

Now we will briefly outline some of the functional analytic properties of the space of bandlimited functions.
We refer the reader to Appendix~\ref{app:fa_sampling} for more details regarding the functional analytic structure of sampling theory.

The sampling theorem for a Nyquist lattice $\{ x_n^{(\alpha)} \}_{n \in \mathds{Z}}$ implies that the collection of $\sinc$ functions centred at the lattice points forms a basis for the space of bandlimited functions.
Moreover, the coefficients of a function in this basis are simply the values of the function at the corresponding lattice points.
This basis is orthonormal in the inner product 
\begin{equation}
(\phi, \psi) := \frac{\Omega}{\pi} \int dx \; \phi^*(x) \psi(x).
\end{equation}
The orthogonality follows from the identity
\begin{equation}
\frac{\Omega}{\pi} \int dx \hspace{1mm} \sinc [ (x-x_n^{(\alpha)}) \Omega ] \sinc [ (x-x_m^{(\alpha)}) \Omega ] = \delta_{nm}.
\end{equation}
Different values of $\alpha$ corresponds to translated versions of the lattice.
The $\sinc$ functions centred at lattice points with a different value of $\alpha$ also form an orthogonal basis. Crucially, however, $\sinc$ functions from different lattices are not orthogonal:
\begin{eqnarray}
\begin{split}
\frac{\Omega}{\pi} \int dx &\hspace{1mm} \sinc [ (x-x_n^{(\alpha)}) \Omega ] \sinc [ (x-x_m^{(\alpha')}) \Omega ] \\
&= \sinc [ (x_n^{(\alpha)} - x_m^{(\alpha')}) \Omega ] \\
&= \sinc [ \pi (n-m) + (\alpha - \alpha')/2 ] \\
&\neq \delta_{nm}, \hspace{3mm} \mbox{for} \hspace{1mm} \alpha \neq \alpha'
\end{split}
\end{eqnarray}
Here $\alpha, \alpha' \in [0,2\pi)$.
This fact will be important when we study the localization of field degrees of freedom.

Let us denote the vector in the function space corresponding to $\sinc[(x-x_n^{(\alpha)}) \Omega]$ by $| x_n^{(\alpha)} )$.
For fixed $\alpha$, we then obtain a resolution of the identity:
\begin{equation}
\sum_{n \in \mathds{Z}} | x_n^{(\alpha)} ) ( x_n^{(\alpha)} | = \mathds{1}
\end{equation}
Taking the union of these bases over $\alpha$, we get an overcomplete basis for the function space with the corresponding $x_n^{(\alpha)}$'s covering $\mathds{R}$.
Thus, we can write down an overcomplete resolution of identity:
\begin{equation}
\label{eqn:ctm_roi}
\frac{\Omega}{\pi} \int_{\mathds{R}} dx \hspace{1mm} | x ) ( x | = \mathds{1},
\end{equation}
where the measure $\Omega dx / \pi$ gives the density of degrees of freedom in space, given by the Nyquist rate.
This overcomplete resolution of the identity is analogous to that for coherent states $\ket{\alpha}$
\begin{equation}
  \frac{1}{\pi} \int_{\mathds{C}} d^2\alpha \hspace{1mm} \ket{\alpha} \bra{\alpha} = \mathds{1},
\end{equation}
where the measure $d^2 \alpha/ \pi = dx dp / 2 \pi \hbar$ gives the density of independent states in phase space.

The space of bandlimited functions is a reproducing kernel Hilbert space \cite{Zayed1993,Higgins1996,Benedetto2001}.
This means that any function in the space can be recovered by means of the reproducing kernel $K(x,x')$ as follows:
\begin{eqnarray}
\psi(x) &=& \int dx' \hspace{1mm} K(x,x') \psi(x'), \nonumber \\
\label{eqn:reprod_kernel}
K(x,x') &:=& \frac{\Omega}{\pi} (x|x') = \frac{\Omega}{\pi} \sinc[(x-x') \Omega ].
\end{eqnarray}

\subsection{Reconstruction formula for quantum fields}

We now apply the sampling theorem to a quantum field $\hat{\phi}$ in 1+1 dimensions, with a Hilbert space $\mathcal{H}$.
Let $\ket{\phi} \in \mathcal{H}$ be an eigenstate of the field operator obeying $\hat{\phi}(x) \ket{\phi} = \phi(x) \ket{\phi}$ $\forall x \in \mathds{R}$ for some real-valued function of eigenvalues, $\phi$.
Note that in a continuum field theory without a cutoff, $\hat \phi$ is not an operator but an operator-valued distribution which must be smeared with a suitable test function to obtain an operator.
In the bandlimited theory, the cutoff acts as an effective smearing function, and $\hat \phi$ is a genuine operator, though still unbounded.

Now let $\mathcal{H}_{(-\Omega,\Omega)}$ be the subspace spanned by the eigenstates of $\hat{\phi}$ where the corresponding functions of eigenvalues $\phi$ are functions bandlimited by $\Omega$.
Because $\phi$ is a bandlimited function, the eigenvalue $\phi(x)$ at any point $x \in \mathds{R}$ can be determined by the knowledge of the eigenvalues $\phi(x_n^{(\alpha)})$ at all of the points $x_n^{(\alpha)}$ on a sampling lattice $\{ x_n^{(\alpha)} \}_{n \in \mathds{Z}}$.
Thus, the action of the operator $\hat{\phi}(x)$ on an eigenstate of the field is determined from its action on a sampling lattice.
Explicitly,
\begin{eqnarray}
\hat{\phi}(x) \ket{\phi} &=& \phi(x) \ket{\phi} \nonumber \\
&=& \sum_{n \in \mathds{Z}} \sinc [(x - x_n^{(\alpha)} ) \Omega] \phi(x_n^{(\alpha)}) \ket{\phi} \nonumber \\
&=& \sum_{n \in \mathds{Z}} \sinc [(x - x_n^{(\alpha)} ) \Omega] \hat{\phi}(x_n^{(\alpha)}) \ket{\phi}.
\end{eqnarray}
Of course, this is true for all eigenstates of $\hat{\phi}$ in $\mathcal{H}_{(-\Omega,\Omega)}$.
Since $\mathcal{H}_{(-\Omega,\Omega)}$ is the span of these eigenstates, the action of $\hat{\phi}(x)$ is determined by that of $\lbrace \hat{\phi}(x_n^{(\alpha)}) \rbrace_{n \in \mathds{Z}}$ for all states in $\mathcal{H}_{(-\Omega,\Omega)}$.
Thus, we can write:
\begin{equation} \label{reconstruction}
\hat{\phi}(x) = \sum_{n \in \mathds{Z}} \sinc [ (x-x_n^{(\alpha)}) \Omega ] \hat{\phi}(x_n^{(\alpha)}).
\end{equation}
Therefore, the operators $\lbrace \hat{\phi}(x_n^{(\alpha)}) \rbrace_{n \in \mathds{Z}}$ form a complete set of commuting observables for any $\alpha$.
The fact that there are many lattices on which the field can be represented means that the localization of the degrees of freedom is non-trivial.
In the next section we will briefly examine this, but it will be studied further in later sections of the paper.

\subsection{A first look at localization}
\label{subsection:firstqm_localization}

In a bandlimited field theory, the local harmonic oscillators, i.e., the local degrees of freedom $\phi(x)$, are not all independent. Namely, a set of degrees of freedom $\{\phi(x_n)\}$ is linearly independent only if the $x_n$ all belong to the same sampling lattice. The field $\phi(x)$ at any other spatial point $x$ can then be reconstructed from the amplitudes $\{\phi(x_n)\}$. 
In this section we will construct a a function that describes the spatial volume occupied by a set of degrees of freedom $\{\phi(x)\}$.

To this end, consider the subspace spanned by $N$ position eigenvectors $\{ |x_n) \}_{n=1}^N$ of first quantization (i.e., they can be represented as number-valued function over space). We will allow that they are not all from the same Nyquist lattice, and therefore they are generally not orthogonal.
Intuitively, the vector $|x_n)$  characterizes the spatial profile of the field degree of freedom $\hat{\phi}(x_n)$ located at this point.
Let us now construct the projector onto this subspace, in the basis of these $N$ position vectors.
First, we map the position vectors to an orthogonal basis
\begin{equation}
|e_i) = \sum_j B_{ij} |x_j).
\end{equation}
The projector onto this subspace is given by
\begin{eqnarray}
\mathds{1}_N &:=& \sum_i |e_i)(e_i| \nonumber \\
&=& \sum_{j,k} \left( \sum_i B_{ij} (B^\dagger)_{ik} \right) |x_j) (x_k|.
\end{eqnarray}
We can express the elements of the projector in the non-orthogonal basis $\{ |x_n) \}_{n=1}^N$ using the reproducing kernel,
\begin{eqnarray}
K(x_j,x_k) &:=& \frac{\Omega}{\pi} ( x_j | x_k ) \nonumber \\
&=& \frac{\Omega}{\pi} \left( \sum_i ( e_i | (B^{-1 \dagger})_{ji} \right) \left( \sum_l (B^{-1})_{kl} | e_l ) \right) \nonumber \\
&=& \frac{\Omega}{\pi} \sum_i (B^{-1 \dagger})_{ji} (B^{-1})_{ki}.
\end{eqnarray}
Viewing $K(x_j,x_k)$ as the $(j,k)^{th}$ element of an $N \times N$ matrix $K_N$, we can write
\begin{equation}
\mathds{1}_N = \frac{\Omega}{\pi} \sum_{j,k} (K_N^{-1})_{kj} |x_j) (x_k|.
\end{equation}

\begin{figure}[ht]
\begin{center}
\includegraphics[height=2.5in,width=3.3in]{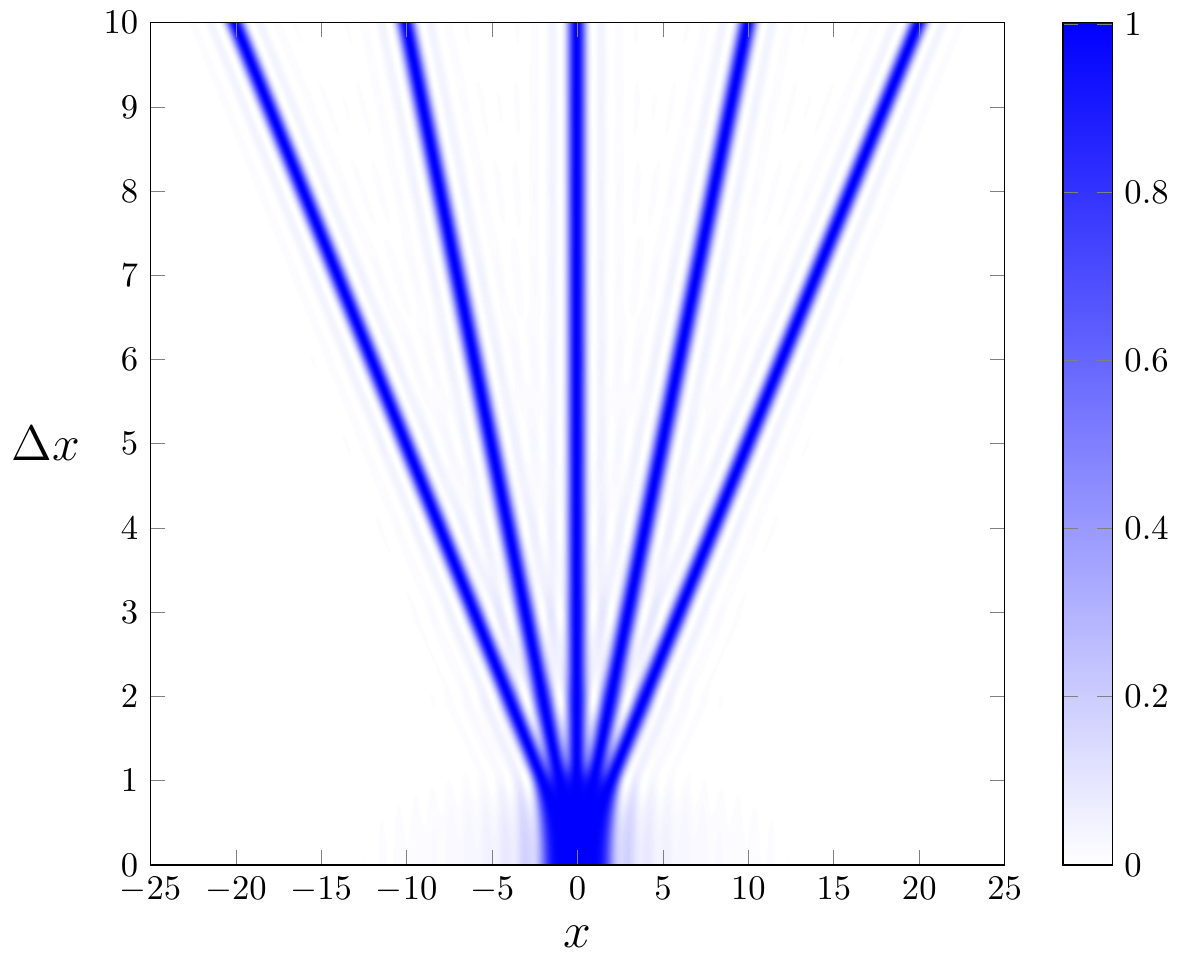}
\caption{\small
  Spatial profile (in the horizontal axis) for 5 degrees of freedom as a function of their spacing (vertical axis).
  Both axes are scaled so that the cutoff length, or Nyquist spacing, $\pi/\Omega$ is equal to 1.
  The sample points are centred at $x=0$.
  We see that for spacings above the Nyquist spacing, the degrees of freedom occupy a region consisting of $5$ disjoint intervals of length $\sim 1$ surrounding each sample point.
  Below the Nyquist spacing, the degrees of freedom merge to occupy a single interval of length $\sim 5$.
  This interval does not decrease in size even as the sampling points are taken on top of one another.
}
\label{plt:spatial_profile}
\end{center}
\end{figure}

Inserting the resolution of identity \eqref{eqn:ctm_roi}, we can write this projector in the continuum basis:
\begin{equation}
\mathds{1}_N = \int_{\mathds{R}^2} dx dx'~ \frac{\Omega}{\pi} \sum_{jk} K(x,x_j) (K_N^{-1})_{kj} K(x_k,x') ~~~ |x) (x'|
\end{equation}
Note that the trace of this operator is $N$, since it is simply the identity on an $N$-dimensional subspace.
This trace is represented in the continuum basis as an integral over the diagonal elements of the integral kernel.
We can interpret the diagonal elements of this projector in the continuum basis (i.e., as a function of $x \in \mathds{R}$) as the spatial profile of the $N$ vectors $\{ |x_n) \}_{n=1}^N$.
We can then visualise the spatial profile of $N$ degrees of freedom for various spacings between the degrees of freedom; this is illustrated in Figure~\ref{plt:spatial_profile}, for 5 points.

We see that for sampling point separations larger than the Nyquist spacing (shown in the region $>1$ on the vertical axis), the degrees of freedom are clearly localized about the chosen sampling points and they are nonlocal on the order of the cutoff length (in the horizontal axis).
When the points are spaced close to the Nyquist spacing (i.e., spacing of 1), the spatial profiles of the points converge to form a volume of length 5.
The surprising point illustrated by this plot is that when the points are spaced closer than a Nyquist spacing apart, the points still describe a spatial volume equivalent to the spatial volume described by the points when they are spaced at the Nyquist spacing.
Thus we see that each point can be localized to a unit of volume, and that these volume elements are incompressible, in the sense that $N$ points pushed close together still describe an interval of length $N$. Intuitively, it is clear that, because of the cutoff, placing the sampling points closer than a Nyquist spacing should not uncover new degrees of freedom. What is surprising but in hindsight plausible is that even when $N$ sample points are closer than the Nyquist spacing they still access $N$ degrees of freedom.  

We will return to this issue in the context of quantum fields in section~\ref{section:localization}.

\subsection{Bandlimited correlation functions}

We would now like to see the effect that bandlimiting a quantum field has on the commutation relations and correlation functions of the ground state of the field.
The particular fields we will hereafter be considering are 1+1 dimensional massless scalar quantum fields with an ultraviolet cutoff $\Omega$ and infrared cutoff $\omega$ imposed on the spatial momentum, so that $\omega < |k| < \Omega$.
Note that bandlimited functions which have support on $(-\Omega,-\omega) \cup (\omega,\Omega)$ in Fourier space form a subspace of the space of functions with support on $(-\Omega,\Omega)$.
Thus the above sampling theorem \eqref{eq:shannon} also applies to these functions.

We consider the system described by the free Klein-Gordon Hamiltonian,
\begin{equation}
H[\phi,\pi] := \frac{1}{2} \int_{\mathds{R}} dx \hspace{1mm} \left( \pi^2(x) + \phi(x) (-\Delta) \phi(x) \right),
\end{equation}
where $\Delta = \partial_x^2$ is the scalar Laplacian operator.
Here and below we will not write the $\hspace{1mm} \hat{} \hspace{1mm}$ on top of operators.
The field $\phi$ and its conjugate momentum $\pi$ can be expressed with the usual mode expansion (see, e.g., \cite{Peskin1995}), except that the field modes outside of the range $\omega < |k| < \Omega$ are removed:
\begin{eqnarray}
\phi(x) &=& \int_{\omega < |k| < \Omega} \frac{dk}{2\pi} \frac{1}{\sqrt{2|k|}} \left( a_k e^{ikx} + a_k^\dagger e^{-ikx} \right), \\
\pi(x) &=& \int_{\omega < |k| < \Omega} \frac{dk}{2\pi} \frac{1}{i} \sqrt{\frac{|k|}{2}} \left( a_k e^{ikx} - a_k^\dagger e^{-ikx} \right).
\end{eqnarray}
Here $a_k, a_k^\dagger$ are the usual annihilation and creation operators obeying the commutation relations $[ a_k, a_{k'}^\dagger ] = (2\pi) \delta (k-k')$ and all other commutators vanish.
Notice that the Hilbert space $\mathcal{H}_{(-\Omega,\Omega)}$ is unitarily preserved under free time evolution because the Fourier modes of the field are uncoupled\footnote{In an interacting theory the modes would couple and, naively, one may expect this to generate shorter-than-cutoff (i.e. transplanckian) wavelengths. Realistically, however, such high-energetic particle collisions would necessarily excite the very gravitational degrees of freedom that are thought to enforce the ultraviolet cutoff in the first place, which may well save unitarity.}.
From the mode expansions we can now calculate commutation relations and two point functions for the field at two arbitrary points $x, x' \in \mathds{R}$.
The equal-time commutation relations between $\phi$ and $\pi$ become
\begin{equation}
\label{eq:commutator}
[ \phi(x), \pi(x') ] = i \left( \frac{\Omega}{\pi} \sinc (\Omega \Delta x) - \frac{\omega}{\pi} \sinc (\omega \Delta x) \right),
\end{equation}
where we have written $\Delta x := x - x'$.
Note that the commutator can be related to the reproducing kernel which we used in the first quantized picture \eqref{eqn:reprod_kernel} after projecting the kernel onto the space of bandlimited functions with support on $(-\Omega,-\omega) \cup (\omega,\Omega)$ in Fourier space:
\begin{equation}
[ \phi(x), \pi(x') ] = i K(x,x').
\end{equation}
The remaining commutators all vanish: $[\phi(x),\phi(x')] = [\pi(x),\pi(x')] = 0$ for any two points.

The correlation function for the ground state of the field at two distinct points is:
\begin{equation} \label{eqn:vac_phicorrlns}
\bra{0} \phi(x) \phi(x') \ket{0} = \frac{1}{2\pi} [ \Ci (\Omega \Delta x) - \Ci (\omega \Delta x) ].
\end{equation}
In the coincidence limit this becomes:
\begin{equation} \label{eqn:vacphi_coincidence}
\bra{0} \phi^2(x) \ket{0} = \frac{1}{2\pi} \log \left( \frac{\Omega}{\omega} \right).
\end{equation}
Similarly, the correlation function for the conjugate momentum at two distinct points is
\begin{equation}
\begin{split}
\label{eqn:vac_picorrlns}
\bra{0} \pi(x) \pi(x') \ket{0} = & \frac{\cos (\Omega \Delta x) - \cos (\omega \Delta x)}{2\pi \Delta x^2} \\& + \frac{\Omega \sin (\Omega \Delta x) - \omega \sin (\omega \Delta x)}{2\pi\Delta x},
\end{split}
\end{equation}
which becomes in the coincidence limit
\begin{equation}
\bra{0} \pi^2 (x) \ket{0} = \frac{\Omega^2 - \omega^2}{4\pi}.
\end{equation}
The correlation functions between $\phi$ and $\pi$ at equal times all vanish, $\frac{1}{2} \bra{0} \lbrace \phi(x), \pi(x') \rbrace \ket{0} = 0$.

From Eq.~\eqref{eqn:vacphi_coincidence}, we see that the infrared cutoff $\omega$ is necessary to regulate the infrared divergence of the $\langle \phi \phi \rangle$ correlator.
It will therefore be necessary to keep $\omega$ strictly positive for our calculations, and we will examine infrared effects in more detail in section \ref{section:ir}.
Since we are more interested in the effect of the ultraviolet cutoff $\Omega$, we display the correlation function in the limit $\omega \to 0$ while adding appropriate counterterms to the correlation functions.
The second term of Eq.~\eqref{eqn:vac_phicorrlns} contains the infrared divergence.
For $\omega/\Omega \ll 1$, we have
\begin{equation}
  \Ci \left( \frac{\omega}{\Omega} \Omega \Delta x \right) = \gamma + \log(\Omega \Delta x) + \log(\omega/\Omega) + \mathcal{O}(\omega/\Omega).
\end{equation}
Therefore if we add a term $\log(\omega/\Omega)/(2\pi)$ to the $\langle \phi \phi \rangle$ correlator, in the limit $\omega \to 0$ we find that at two distinct points
\begin{equation}
  \bra{0} \phi(x) \phi(x') \ket{0} \mapsto \frac{1}{2\pi} [ \Ci(\Omega \Delta x) - \gamma - \log(\Omega \Delta x) ].
\end{equation}
We also want this correlator to give the usual nonbandlimited correlator when $\Omega \to \infty$.
The nonbandlimited correlator is
\begin{equation}
  \bra{0} \phi(x) \phi(x') \ket{0} = \int_{\mathds{R}} \frac{dk}{2\pi} \hspace{1mm} \frac{1}{2|k|} e^{ik\Delta x} = \frac{-1}{2\pi} ( \gamma + \log|\Delta x| ).
\end{equation}
Thus, since $\Ci(\Omega |\Delta x|) \to 0$ as $\Omega \to \infty$, we must add another factor of $\log(\Omega)/(2\pi)$ to $\langle \phi \phi \rangle$ to get
\begin{equation}
  \bra{0} \phi(x) \phi(x') \ket{0} = \frac{1}{2\pi} [ \Ci(\Omega \Delta x) - \gamma - \log|\Delta x| ],
\end{equation}
and at a single point we have
\begin{equation}
  \bra{0} \phi^2(x) \ket{0} = \frac{1}{2\pi} \log(\Omega).
\end{equation}
The correlator $\langle \pi \pi \rangle$ is not divergent in the infrared, thus we can directly take the limit $\omega/\Omega \to 0$.

\begin{figure}[ht]
\begin{center}
\includegraphics[height=2.5in,width=3.3in]{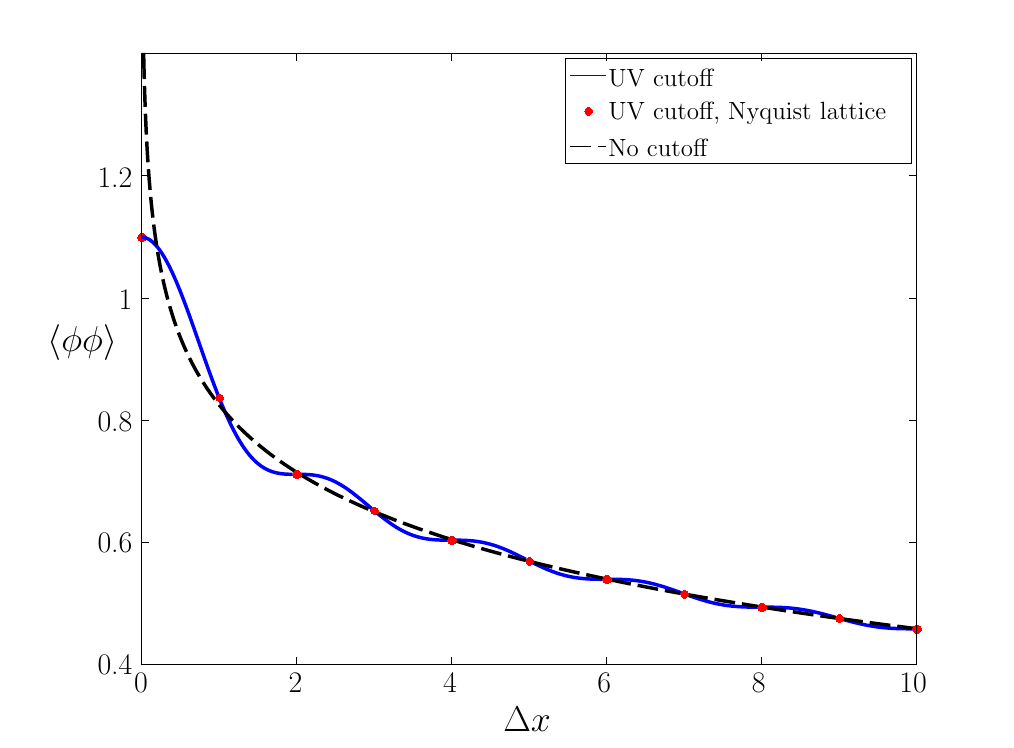}
\caption{\small
$\phi$-$\phi$ correlations as a function of their separation.
The horizontal axis is scaled by $\Omega/\pi$ so that integer values correspond to Nyquist spacings.
The bandlimited correlations are blue with the Nyquist spacings indicated by red dots.
The black dashed line shows the ultraviolet-divergent correlations without the ultraviolet bandlimit.
We see that for points on the Nyquist lattice, the bandlimited correlators are in closer agreement to the correlation functions without the bandlimit.
A counterterm of $\log(\omega)/(2\pi)$ is added to $\langle \phi \phi \rangle$ to cancel the infrared divergence.
}
\label{plt:phi_corrlns}
\end{center}
\end{figure}

\begin{figure}[ht]
\begin{center}
\includegraphics[height=2.5in,width=3.3in]{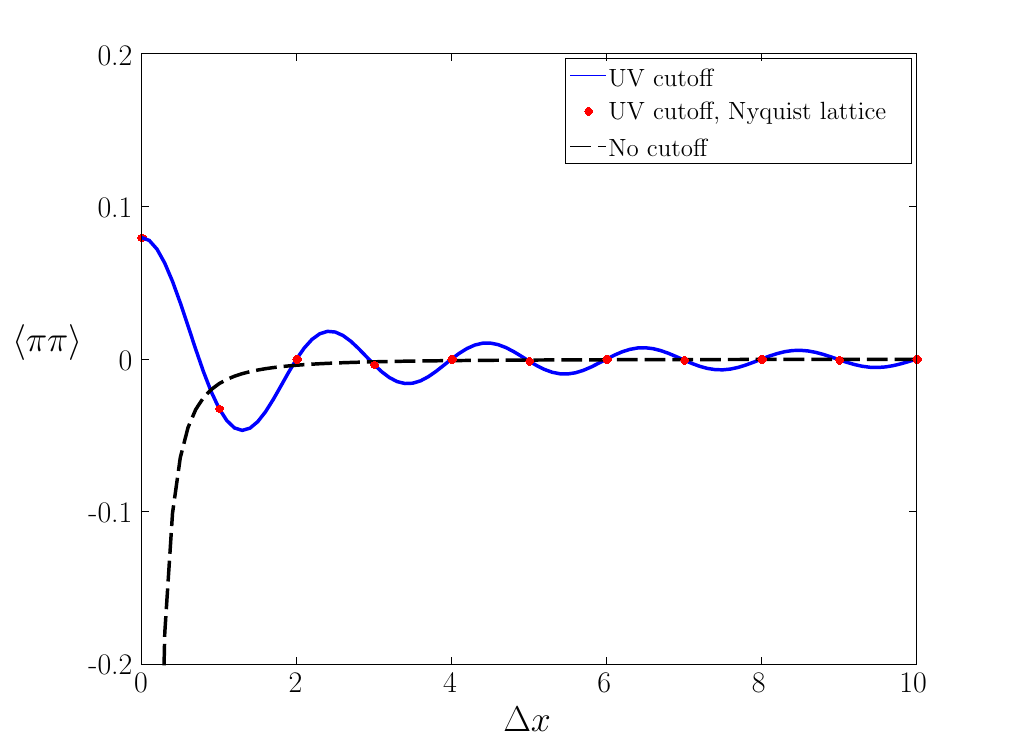}
\caption{\small
  $\pi$-$\pi$ correlations as a function of their separation.
  The horizontal axis is scaled by $\Omega/\pi$ so that integer values correspond to Nyquist spacings.
  The vertical axis is scaled by $1/\Omega^2$.
  The bandlimited correlations are blue with the Nyquist spacings indicated by red dots.
  The black dashed line shows the ultraviolet-divergent correlations without the ultraviolet bandlimit.
}
\label{plt:pi_corrlns}
\end{center}
\end{figure}

In Figures \ref{plt:phi_corrlns} and \ref{plt:pi_corrlns} we present graphs of the correlation functions $\bra{0} \phi(x) \phi(x') \ket{0}$ and $\bra{0} \pi(x) \pi(x') \ket{0}$ (respectively) as a function of $\Delta x$, with the infrared cutoff removed.
In both figures, we see that the correlations decay with distance, similarly to the correlations when there is no ultraviolet cutoff.
We notice that the bandlimited correlation functions oscillate with wavelengths of the order of the ultraviolet cutoff length.
As we will see later, similar oscillations also occur in the correlations of thermally distributed bandlimited classical signals.

\section{Entropy of Gaussian states}
\label{section:gaussian}

We would like to understand how the delocalization of degrees of freedom of the bandlimited quantum field theory affects the localization of information in space.
A useful tool to probe the distribution of degrees of freedom in space is the von Neumann entropy of localized subsystems, and derived quantities such as the mutual information.
Suppose we have a state of a bandlimited field, whose wavenumbers are in the range $\omega < \abs{k} < \Omega$, and we are able to make measurements of the field amplitude $\phi$ and its conjugate momentum $\pi$ at a finite number, $N$, of points.
Any set of points defines a subsystem in this way, regardless of their relative positions in space, and we can define a von Neumann entropy associated with this subsystem.
In this section we show how to calculate the von Neumann entropy associated with an arbitrary subset of points using the formalism of Gaussian states \cite{Adesso2007}.
We leave concrete numerical calculations for section \ref{section:ee}.

First in section \ref{subsection:classical}, we consider a classical field in a thermal state.
For this system there is no entanglement or quantum noise and all of the entropy comes from thermal fluctuations. 
Then in \ref{subsection:quantum}, we consider a thermal state of a quantum system, which in the limit of zero temperature becomes the ground state of the quantum field. 
In this case the entropy includes both entanglement entropy and thermal fluctuations, and in the zero temperature limit is entirely due to quantum entanglement.
This will demonstrate how the classical results emerge from the quantum results at high temperatures.

\subsection{Classical entropy}
\label{subsection:classical}

Let us first consider the classical situation where the field we are given is chosen from some given classical distribution.
If, for example, the statistical distribution is that of Gaussian white noise, the values of the field at different sample points are uncorrelated.
Since the entropy is additive for uncorrelated degrees of freedom, the entropy associated with $N$ samples is linear in $N$.
The classical case which we will be more interested in, because it is more comparable to the quantum case, is the case where the statistical distribution of the signals is a thermal distribution of a 1+1 dimensional classical Klein-Gordon Hamiltonian, with fields that have both an ultraviolet cutoff $\Omega$ and infrared cutoff $\omega$.
For this distribution, the degrees of freedom of the field at the sample points will be coupled, causing the values of the field at these points to be correlated.
We therefore expect that the entropy of the interval of $N$ samples has a contribution which is nonlinearly dependent on $N$.
We now explicitly calculate the entropy created for these $N$ samples.

We will go through the entropy calculation in detail since many of the steps are reproduced in the quantum calculation.
We begin with the Hamiltonian in momentum space,
\begin{equation}
H [ \tilde{\phi}, \tilde{\pi} ] = \frac{1}{2} \int_{\omega < \abs{k} < \Omega} \frac{dk}{2\pi} \hspace{1mm} \left( | \tilde{\pi}(k) |^2 + k^2 | \tilde{\phi}(k) |^2 \right).
\end{equation}
Both the ultraviolet and infrared cutoffs are explicitly enforced in this description by the restriction on the momentum $k$.
For a thermal state, the probability distribution of the fields in momentum space is given by the Boltzmann distribution
\begin{equation}
\label{eq:class_pdf_momm}
p[\tilde{\phi},\tilde{\pi}] = \frac{1}{Z} e^{-\beta H[\tilde{\phi},\tilde{\pi}]}.
\end{equation}

We seek the entropy of a reduced probability distribution for a set of $N$ sample points.
To this end, we choose a lattice $\lbrace x_n \rbrace_{n \in \mathds{Z}}$ and we write $\Delta x_{mn} := x_m - x_n$.
We denote the values of the field at these points $\phi_n := \phi(x_n)$, $\pi_n := \pi(x_n)$.
To find the probability distribution for the samples, we perform a change of phase space coordinates from the fields in momentum space to a sampling lattice:
\begin{eqnarray}
\label{eq:cov_momm_lattice}
\phi (x_n) &=& \int_{\omega < |k| < \Omega} \frac{dk}{2 \pi} e^{i k x_n} \tilde \phi(k), \nonumber \\
\pi (x_n) &=& \int_{\omega < |k| < \Omega} \frac{dk}{2 \pi} e^{i k x_n} \tilde \pi(k).
\end{eqnarray}
Instead of explicitly performing this change of variables, it will be more convenient to make use of the fact that the probability distribution in terms of $\phi_n$ and $\pi_n$ is Gaussian.
This follows from the fact that the probability distribution in momentum space is Gaussian and the change of phase space variables \eqref{eq:cov_momm_lattice} is linear.
The distribution is therefore characterized entirely by the two-point functions.

First, note that one can easily calculate the power spectra of the fields from the probability distribution \eqref{eq:class_pdf_momm} where the fields are represented in momentum space:
\begin{eqnarray}
\label{eq:classphiflucs}
\langle | \tilde{\phi}(k) |^2 \rangle &=& \frac{1}{\beta \omega_k^2}, \\
\label{eq:classpiflucs}
\langle | \tilde{\pi}(k) |^2 \rangle &=& \frac{1}{\beta}.
\end{eqnarray}
Then we can obtain the correlators of the fields between two arbitrary points by taking the Fourier transform of the power spectra:
\begin{eqnarray}
\label{eq:classphicorrlns}
\begin{split}
\langle \phi(x) \phi(x') \rangle = & \frac{1}{\beta} \left[ \frac{\cos (\omega \Delta x)}{\pi \omega} - \frac{\cos (\Omega \Delta x)}{\pi \Omega} \right] \\
&+ \frac{\Delta x}{\beta} \left[ \Si(\omega \Delta x) - \Si(\Omega \Delta x) \right],
\end{split} \\
\label{eq:classpicorrlns}
\langle \pi(x) \pi(x') \rangle = \frac{1}{\beta} \left[ \frac{\Omega}{\pi} \sinc (\Omega \Delta x) - \frac{\omega}{\pi} \sinc (\omega \Delta x) \right].
\end{eqnarray}

These correlators are plotted as a function of $\Delta x$ in Figures~\ref{plt:hightemp_phicorrlns} and \ref{plt:hightemp_picorrlns} alongside the finite-temperature quantum correlators calculated below.
In a similar manner as for the vacuum correlators in the previous section, these functions are displayed with added infrared counterterms.
The $\langle \pi \pi \rangle$ correlator is again finite as $\omega/\Omega \to 0$, so for plotting purposes, we can just take this limit.
Since $\Si(\omega \Delta x) \to 0$ as $\omega/\Omega \to 0$, the first term in Eq.~\eqref{eq:classphicorrlns} is the only infrared divergent term, whose divergence may be cancelled by subtracting $1/(\pi \beta \omega)$ since as $\omega/\Omega \to 0$,
\begin{equation}
  \frac{\cos(\omega \Delta x)}{\pi \beta \omega} \sim \frac{1}{\pi \beta \omega}.
\end{equation}
The function plotted in figure~\ref{plt:hightemp_phicorrlns} is then
\begin{equation}
  \langle \phi(x) \phi(x') \rangle = \frac{-1}{\beta} \left[ \frac{\cos (\Omega \Delta x)}{\pi \Omega} + \Delta x \Si(\Omega \Delta x) \right].
\end{equation}

For the entropy calculation, we require the reduced probability distribution for $N$ sample points obtained by marginalizing over a complementary set of phase space coordinates.
In a lattice theory, this would simply mean integrating over the field values at all lattice sites not under consideration.
In a bandlimited theory, the complementary subsystem to a set of points is not generally associated to any set of points; it is nonlocal.
As previously noted, we do not actually need to carry out the marginalization, as the reduced probability distribution on the $N$ samples is Gaussian, and hence entirely determined by the two-point functions of the $N$ points.
Nevertheless it is instructive to carry out the decomposition of the total system into the $N$ sample points and their complement.

We will perform this calculation for any Gaussian state of the form
\begin{equation}
\begin{split}
  p( \{ \phi_n, \pi_n \}_n ) =
\frac{1}{Z} & e^{-\tfrac12 \sum_{m,n} \pi_m A_{mn} \pi_n} \\
  &\times e^{-\tfrac12 \sum_{m,n} \phi_m B_{mn} \phi_n}
\end{split}
\end{equation}
with positive-definite, symmetric matrices $A$ and $B$.
For the above example, the $\phi_n$'s and $\pi_n$'s would represent the field amplitudes on a sampling lattice.
We can split the phase space into a subsystem describing the $N$ sample points and its complementary subsystem by performing a change of variables which splits the matrices encoding the Poisson bracket and symplectic form into a direct sum of the matrices which act on the individual subspaces separately.
It will be convenient to write the phase space variables in a vector $\vec{r} = (\phi_1, \pi_1, \phi_2, \pi_2, \ldots)$.
Then the Poisson bracket for the total phase space with the ultraviolet and infrared cutoff constraints enforced can be encoded in an anti-symmetric matrix $\Lambda_{ij} := \lbrace r_i, r_j \rbrace_{PB}$, where
\begin{eqnarray}
\label{eq:Dirac_bracket}
\lbrace \phi_i, \pi_j \rbrace_{PB} &=& \frac{\Omega}{\pi} \sinc (\Omega \Delta x_{ij}) - \frac{\omega}{\pi} \sinc (\omega \Delta x_{ij}), \nonumber \\
\lbrace \phi_i, \phi_j \rbrace_{PB} &=& \lbrace \pi_i, \pi_j \rbrace_{PB} = 0.
\end{eqnarray}
Note that this consistent with the quantum mechanical commutation relations \eqref{eq:commutator}.

First we write the Poisson bracket as a block matrix,
\begin{equation} \label{Lambda}
\Lambda = \left[
\begin{array}{cc}
 \alpha & \eta \\
 -\eta^T & \gamma 
\end{array}
\right],
\end{equation}
where $\alpha^T = -\alpha$ and $\gamma^T = -\gamma$, ensuring that $\Lambda^T = -\Lambda$.
The blocks are arranged such that the upper left block contains the first $2N$ indices which correspond to the $N$ sample points which will remain after marginalizing.
Now consider a change of phase space variables $\vec{r'} = Q \vec{r}$, where
\begin{equation}
\label{eq:phase_split}
Q := \left[
\begin{array}{cc}
 I & 0 \\
 \eta^T \alpha^{-1} & I
\end{array}
\right].
\end{equation}
The transformed Poisson bracket is
\begin{equation}
\Lambda' := Q \Lambda Q^T = \left[
\begin{array}{cc}
 \alpha & 0 \\
 0 & \eta^T \alpha^{-1} \eta + \gamma
\end{array}
\right].
\end{equation}
Now the phase space splits as a direct sum of the $N$ degrees of freedom remaining after marginalizing and its complement, with symplectic form
\begin{equation}
\label{eq:subspace_vol}
\begin{split}
F := & \sum_{n,m =1}^N (\alpha^{-1})_{nm} d\phi'_n \wedge d\pi'_n\\
&+ \sum_{n,m \not\in \{1,\dots,N\} } [(\eta^T \alpha^{-1} \eta + \gamma)^{-1}]_{nm} d\phi'_n \wedge d\pi'_n.
\end{split}
\end{equation}

This construction requires that the matrix $\alpha$ be invertible, but we will now show that in the bandlimited theory this is always the case.
We can express $\alpha$ as the matrix $\alpha_{ij} = K(x_i,x_j) = \frac{\Omega}{\pi} (x_i|x_j)$, where $|x_i),|x_j)$ are the bandlimited position eigenfunctions defined in \ref{subsection:firstqm_localization}.
Suppose that $\alpha$ is not invertible.
Then we can find a vector $a_i$ such that 
\begin{equation}
0 = a_i \alpha_{ij} a_j = \sum_{i,j} (x_i| a_i a_j |x_j) = \Vert \sum_i a_i |x_i) \Vert^2,
\end{equation}
and hence $\sum_i a_i |x_i) = 0$.
This implies that for any bandlimited function $f$, we have
\begin{equation} \label{f}
0 = (f| \sum_i a_i |x_i) = \sum_i a_i (f|x_i) = \sum_i a_i f^\ast(x_i).
\end{equation}
However given any finite set of points $x_i$ and target values $y_i$, we can find a bandlimited function $f$ such that $f(x_i) = y_i$ \cite{Kempf:1999tq}.
This contradicts \eqref{f}, hence $\alpha$ must be invertible.

Now, the probability distribution after performing the marginalization is of the form:
\begin{equation}
\begin{split}
p_{\mbox{red}}( \{ \phi_n, \pi_n \}_{n=1}^N ) =
\frac{1}{Z_{\mbox{red}}} & e^{-\frac{1}{2} \sum_{m,n =1}^N \pi_m (A|_N)_{mn} \pi_n} \\
&\times e^{-\frac{1}{2} \sum_{m,n =1}^N \phi_m (B|_N)_{mn} \phi_n}
\end{split},
\end{equation}
where $A|_N$ and $B|_N$ are positive-definite symmetric matrices given by
\begin{eqnarray}
\label{eq:lattice_phicorrlns}
(B|_N^{-1})_{mn} &=& \langle \phi_m \phi_n \rangle,\\
\label{eq:lattice_picorrlns}
(A|_N^{-1})_{mn} &=& \langle \pi_m \pi_n \rangle.
\end{eqnarray}
These correlators are just \eqref{eq:classphicorrlns} and \eqref{eq:classpicorrlns} evaluated at the points $x = x_m$ and $x' = x_n$.
Note that the matrices $A|_N$ and $B|_N$ are simply the original matrices $A$ and $B$ with indices restricted to the subsystem $m,n \in \{1,\dots,N\}$.

Before we can calculate the partition function by integrating over the phase space variables $\{ \phi_n, \pi_n \}_{n=1}^N$, we must first choose a measure for this phase space.
When calculating the entropy for a classical probability distribution with continuous random variables, there is an ambiguity in the entropy caused by an ambiguity in the choice of measure.
Here, we will fix the measure to be proportional to the symplectic form in this subspace (Eq.~\eqref{eq:subspace_vol}) $d \phi_n \wedge d \pi_n / (2 \pi)$.
This is sufficient to resolve the ambiguity in the entropy, and ensures that the entropy invariant under symplectic transformations of the phase space.
As we shall see, the factor of $2 \pi$ gives an entropy which matches the high temperature limit of the von Neumann entropy in the quantum setting.

\begin{figure}[ht]
\begin{center}
\includegraphics[height=2.5in,width=3.3in]{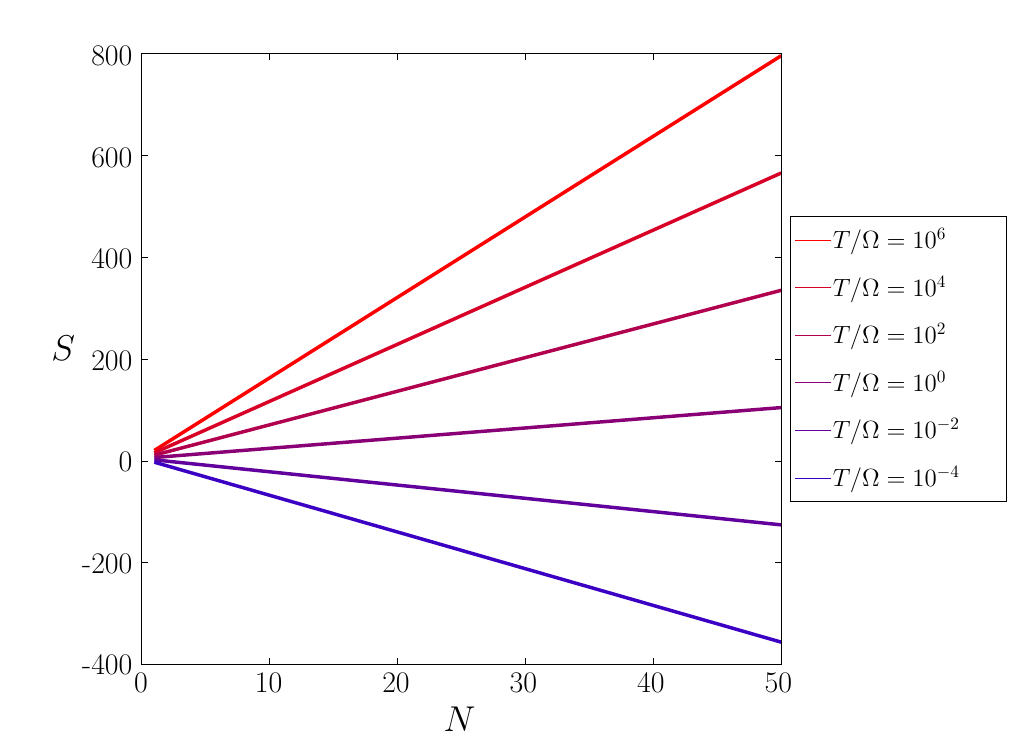}
\caption{\small
Entropy of a set of sample points for a thermally-distributed classical bandlimited Klein-Gordon field for several temperatures.
We see that the magnitude of the entropy grows linearly with the number of points, illustrating a volume law.
Notice also that this figure shows negative entropy at low temperatures, which simply reflects the fact that continuous probability distributions can have negative entropy.
We fix the inherent ambiguity in these entropies by requiring that it matches the quantum entropy at high temperature.
}
\label{plt:Scl_thermal}
\end{center}
\end{figure}

Now we find that the reduced partition function is
\begin{eqnarray}
Z_{\mbox{red}} &=& \int \det(\alpha)^{-1} \prod_{n=1}^N \frac{d\phi_n d\pi_n}{2\pi} \hspace{1mm}
\begin{split}
&e^{-\frac{1}{2} \sum_{m,n =1}^N \pi_m A_{mn} \pi_n} \\
&\times e^{-\frac{1}{2} \sum_{m,n =1}^N \phi_m B_{mn} \phi_n}
\end{split} \nonumber \\
&=& \det(\alpha)^{-1} \sqrt{ \det \left( A^{-1} \right) \det \left( B^{-1} \right) }.
\end{eqnarray}
Then the entropy of the distribution of the remaining samples is:
\begin{eqnarray}
S &=& \int \det(\alpha)^{-1} \prod_{n=1}^N \frac{d\phi_n d\pi_n}{2\pi} \hspace{1mm} p_{\mbox{red}}( \{ \phi_n, \pi_n \}_{n=1}^N ) \nonumber \\
&& \hspace{2cm} \times \log \left[ p_{\mbox{red}}( \{ \phi_n, \pi_n \}_{n=1}^N ) \right] \nonumber \\
&=& N + \log \left[ \det(\alpha)^{-1} \sqrt{ \det \left( A^{-1} \right) \det \left( B^{-1} \right) } \right].
\end{eqnarray}
Note that the first term is linear in the number of samples $N$, but the second term, which encodes the coupling between the sample points, is not exactly linear.
In Figure~\ref{plt:Scl_thermal}, we plot the entropy for a massless bandlimited field in a thermal state.
We find that the entropy is very close to linear in the number of lattice points, regardless of the temperature.
This simply reflects the extensivity of thermal entropy.

It turns out to be more convenient to rewrite the formula for the entropy in terms of the following matrices:
\begin{align}
\Lambda|_N &:= \left[
\begin{array}{cc}
 0 & \lambda \\
 -\lambda & 0
\end{array}
\right], \\
\Sigma|_N &:= \left[
\begin{array}{cc}
 B|_N^{-1} & 0 \\
 0 & A|_N^{-1}
\end{array}
\right],
\end{align}
where $|_N$ denotes restriction of the matrix indices to $1, \dots, N$ (i.e. the indices corresponding to the remaining sample points after marginalizing).
The matrix $\Lambda|_N$ encodes the Poisson bracket with the indices of the phase space vector $\vec{r}$ rearranged so that all of the $\phi_n$'s occur before the $\pi_n$'s, and where $\lambda_{mn} := \{ \phi_m, \pi_n \}_{PB}$. 
Note that $\lambda$ is just the matrix $\alpha$ from Eq.~\eqref{Lambda} after a permutation of the indices.
Similarly, the matrix $\Sigma|_N$ is simply the matrix of correlators between the phase space variables.

It is then straightforward to verify that the entropy can be written as
\begin{eqnarray}
S &=& N + \log \left[ \sqrt{\det(\Lambda|_N^{-1} \Sigma|_N)} \right] \nonumber \\
\label{eq:classical_entropy}
&=& \sum_{i = 1}^N ( 1 + \log(d_i) ),
\end{eqnarray}
where $\{ d_i \}_{i=1}^N$ are the positive imaginary parts of the eigenvalues of $\Lambda|_N^{-1} \Sigma|_N$ which come in pairs $\pm i d_i$.
Therefore, the entropy is determined entirely in terms of the Poisson bracket \eqref{eq:Dirac_bracket} and two-point correlators \eqref{eq:classphicorrlns} and \eqref{eq:classpicorrlns}.
Notice that since we can write down the bracket and correlators between any two points of the field (not necessarily on a Nyquist lattice), we can calculate the entropy associated with $N$ arbitrary points of the field.

\subsection{Quantum entropy}
\label{subsection:quantum}

Let us now proceed to the case of calculating the corresponding entropy for $N$ samples of a quantum field.
Here we will review a method for calculating the entanglement entropy for Gaussian states of systems of quantum harmonic oscillators.
The Gaussian formalism for calculating the entropy for Gaussian states goes back to \cite{Sorkin1983,Bombelli1986,Srednicki1993} with significant simplifications developed in \cite{Audenaert2002,Adesso2007}.
Crucial for our purposes is that this formalism also generalizes naturally to the case where the commutation relations between $\phi$ and $\pi$ are nonlocal.
As in the classical case, the state of a region, and hence its entropy, is determined entirely by the matrix of Poisson brackets and the covariance matrix.

We begin with the Hamiltonian for a set of $M$ coupled oscillators (where $M$ may be finite or infinite):
\begin{equation}
H := \frac{1}{2} \sum_{m,n=1}^M \pi_m A_{mn} \pi_n + \frac{1}{2} \sum_{m,n=1}^M \phi_m B_{mn} \phi_n.
\end{equation}
Here $A$ and $B$ are positive-definite, symmetric matrices.
As in section \ref{subsection:classical} it will be convenient to combine $\phi_n$ and $\pi_n$ into a single vector $\vec{r} = (\phi_1, \pi_1, \phi_2, \pi_2, \ldots)$.
Then we define the correlation matrix
\begin{equation}
\label{eq:covmtx}
\Sigma_{ij} := \tfrac12 \langle \{ r_i, r_j \} \rangle \equiv \tfrac12 \tr \left( \{ r_i, r_j \} \rho \right),
\end{equation}
where $\rho$ is the density matrix for the state.
Also, since the commutators are c-numbers, we can define the matrix $\Lambda$ that encodes the commutation relations:
\begin{equation}
i \Lambda_{ij} \mathds{1} := [r_i, r_j].
\end{equation}

Now we would like to find the state of a subset of oscillators, $i = 1, \ldots, N$ (with $N < M$) obtained by tracing over the complementary set of degrees of freedom.
For the bandlimited theory, the complementary set of degrees of freedom are not simply the complementary set of lattce points $N+1,\dots,M$.
Notice that if we remove the infrared cutoff, i.e., set $\omega = 0$, then on a Nyquist lattice we have the canonical commutation relations $[\phi(x_m), \pi(x_n)] = i \frac{\Omega}{\pi} \delta_{mn}$.
Therefore, on this Nyquist lattice the Hilbert space $\mathcal{H}_{(-\Omega,\Omega)}$ factors into the tensor product $\otimes_n \mathcal{H}_{x_n}$ of the Hilbert spaces $\mathcal{H}_{x_n}$ generated by the field operators $\lbrace \phi(x_n), \pi(x_n) \rbrace$.
However, for a general sampling lattice the commutation relations are nonlocal at the cutoff scale.
Thus, in the case of non-Nyquist sampling, the Hilbert space $\mathcal{H}_{(-\Omega,\Omega)}$ does not simply factor into a tensor product of Hilbert spaces generated by the field operators at the sample points; i.e., $\mathcal{H}_{(-\Omega,\Omega)} \neq \otimes_n \mathcal{H}_{x_n}$.
This is because for points, say $x_n$ and $x_m$, which do not lie on the same Nyquist lattice, the operator $\phi(x_n)$ acts non-trivially on $\mathcal{H}_{x_m}$ since this sector of the Hilbert space is generated by $\phi(x_m)$ and $\pi(x_m)$, but $[ \phi(x_n), \pi(x_m) ] \neq 0$.
However, if we identify a finite subset of lattice points it is possible to factor the Hilbert space into a tensor product of a subspace describing the subsystem and its complement.

Factoring the Hilbert space is similar to the phase space splitting performed in the classical case.
We proceed using the change of variables defined by the matrix \eqref{eq:phase_split}, so that $\Lambda$ takes the form:
\begin{equation}
\Lambda' := Q \Lambda Q^T = \left[
\begin{array}{cc}
 \alpha & 0 \\
 0 & \eta^T \alpha^{-1} \eta + \gamma
\end{array}
\right].
\end{equation}
Note that the block $\alpha := \Lambda|_J$ corresponding to the degrees of freedom $\{ \phi_i, \pi_i \}_{i=1}^N$ is unchanged under this transformation.
The covariance matrix also transforms as
\begin{equation}
\Sigma' := Q \Sigma Q^T.
\end{equation}
It is also easy to check that the block $\Sigma|_J$ is unchanged under this transformation.

Now, by Darboux's theorem, we can find a change of variables via a transformation $T$ which brings the matrix $\Lambda'$ into its canonical form.
That is, we can find a set of canonical coordinates satisfying canonical commutation relations $[ q_i, p_j ] = i \delta_{ij}$.
Since we have split the phase space into two pieces, we can find such a matrix $T$ in the form
\begin{equation}
T = \left[
\begin{array}{cc}
 T_1 & 0 \\
 0 & T_2
\end{array}
\right].
\end{equation}
Now,
\begin{eqnarray}
\Lambda'' &:=& T \Lambda' T^T = \bigoplus_{i=1}^M \left[ \begin{array}{cc}
 0 & 1 \\
 -1 & 0
\end{array}\right] \\
\Sigma'' &:=& T \Sigma' T^T.
\end{eqnarray}
An advantage of working with covariance matrices is that the tracing operation can be implemented by simply restricting the indices of $\Lambda''$ and $\Sigma''$ to the set $\{ 1,\ldots,2N \}$.
Also, by construction of the matrices $Q$ and $T$, the reduced commutator and covariance matrices can be determined solely from the reduced commutator and covariance matrices of the original variables, i.e.,
\begin{eqnarray}
\Lambda''|_N &=& T_1 \Lambda|_N T_1^T \\
\Sigma''|_N &=& T_1 \Sigma|_N T_1^T.
\end{eqnarray}
Therefore, we see that since we are only mixing the coordinates for the degrees of freedom $1, \ldots, N$ among themselves before the tracing operation, the Hilbert space after the tracing operation is the same as the Hilbert space generated by the original operators $\{ \phi_n, \pi_n \}_{n =1}^N$.
The importance of this fact is that it does not matter where the remainder of the samples are taken, as we can simply identify the entire Hilbert space $\mathcal{H}_{(-\Omega,\Omega)}$ as the tensor product of the Hilbert space generated by the operators corresponding to $\{ \phi_n, \pi_n \}_{n=1}^N$ and the Hilbert space of the complementary set of degrees of freedom.
This also allows us to choose a set of samples which do not all lie on a Nyquist lattice, as we can simply identify the reduced Hilbert space as the space generated by the operators at these points.

Now that we have the reduced commutator and covariance matrices in a Darboux basis, as shown by Williamson \cite{Williamson1936,Williamson1937}, we can make a further (symplectic) transformation $S$ that preserves the form of $\Lambda''|_N$, and puts $\Sigma''|_N$ into diagonal form:
\begin{eqnarray}
S \Lambda''|_N S^T &=& \Lambda''|_N, \\
S \Sigma''|_N S^T &=& \bigoplus_{i=1}^N \left[ \begin{array}{cc} d_i & 0 \\ 0 & d_i \end{array} \right].
\end{eqnarray}
Here the diagonal entries $d_i$ come in pairs, and are called the symplectic eigenvalues of $\Sigma''|_N$.

In these coordinates, the density matrix is a product of uncorrelated thermal density matrices of $N$ harmonic oscillators with canonical coordinates $q_i$, $p_i$ satisfying canonical commutation relations, in a state where $\avg{q_i}^2 = \avg{p_i}^2 = d_i$.
In particular, the $i^\text{th}$ oscillator is in the state
\begin{equation}
\rho_i = \sum_{n \geq 0} \frac{1}{d_i + \tfrac12} \left( \frac{d_i - \tfrac12}{d_i + \tfrac12} \right)^n |n\rangle \langle n |.
\end{equation}
Note that the uncertainty relation implies
\begin{equation}
d_i = \Delta q_i \Delta p_i \geq \tfrac12 \abs{ [q_i,p_i] } = \tfrac12.
\end{equation}
Thus the symplectic eigenvalues are all bounded below by $d_i \geq \tfrac12$.

The entropy is then the sum of entropies of each individual oscillator:
\begin{equation}
\begin{split}
S &= \sum_i S(d_i), \\
S(d) &:= (d + \tfrac12) \log (d + \tfrac12) - (d - \tfrac12) \log (d - \tfrac12).
\end{split}
\end{equation}
Thus we can determine the entropy entirely from the symplectic eigenvalues $d_i$.

Note that one does not need to carry out this symplectic diagonalization in order to find the symplectic eigenvalues.
Under the sequence of similarity transformations $Q$, $T$, $S$, the eigenvalues of $\Lambda|_N^{-1} \Sigma|_N$ are invariant, i.e.,
\begin{equation}
\begin{split}
\spec & \left( \Lambda''|_N^{-1} \bigoplus_{i=1}^N \left[ \begin{array}{cc} d_i & 0 \\ 0 & d_i \end{array} \right] \right) \\
&= \spec \left( [ (ST_1) \Lambda|_N (ST_1)^T ]^{-1} (ST_1) \Sigma|_N (ST_1)^T \right) \\
&= \spec \left( \Lambda|_N^{-1} \Sigma|_N \right),
\end{split}
\end{equation}
where $\spec(A)$ denotes the spectrum of $A$.
We therefore see that the eigenvalues of $\Lambda|_N^{-1} \Sigma|_N$ coincide with the eigenvalues of the matrix
\begin{equation}
\Lambda''|_N^{-1} \bigoplus_{i=1}^N \left[ \begin{array}{cc} d_i & 0 \\ 0 & d_i \end{array} \right] = \bigoplus_{i=1}^N \left[ \begin{array}{cc} 0 & d_i \\ -d_i & 0 \end{array} \right],
\end{equation}
which are the values $\pm i d_i$.
Thus we can find the symplectic eigenvalues simply by finding the eigenvalues of $\Lambda|_N^{-1} \Sigma|_N$.
See \cite{Adesso2007}, as well as the related result \cite{Sorkin2012}.

\begin{figure}[ht]
\begin{center}
\includegraphics[height=2.5in,width=3.3in]{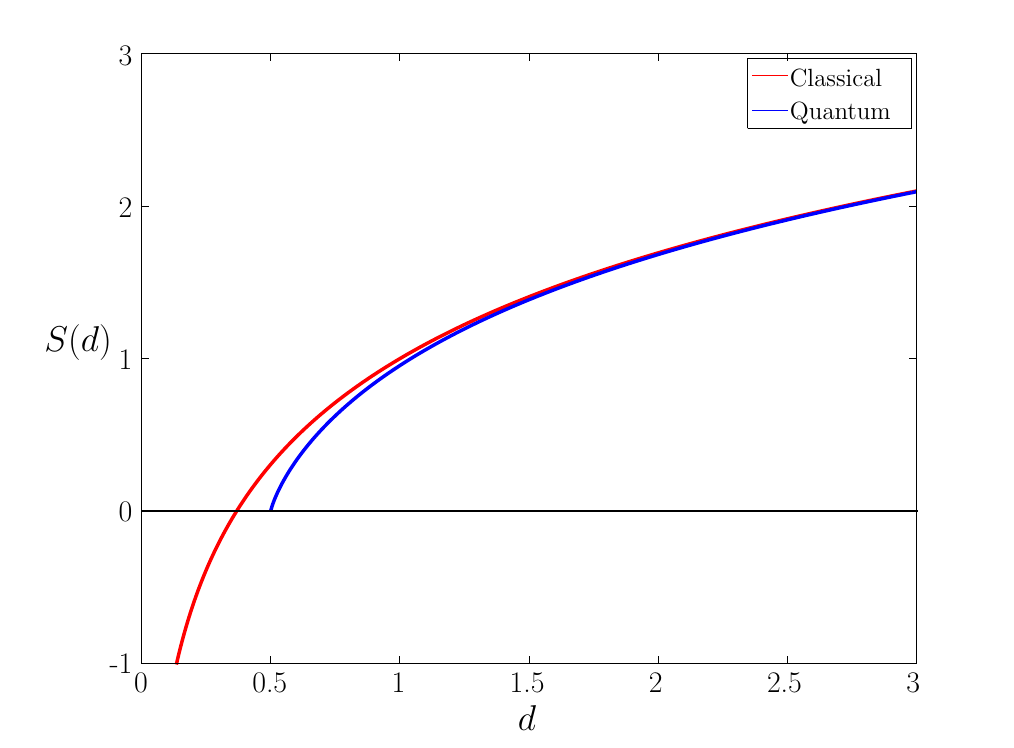}
\caption{\small The entropy of a harmonic oscillator as a function of the symplectic eigenvalue, both classically and quantum-mechanically. In quantum mechanics, the uncertainty principle requires $d \geq 1/2$, which is saturated by the vacuum state for which the entropy is zero. 
Classically, the symplectic eigenvalue can be any positive number, but the entropy becomes negative for small $d$. 
This again is a consequence of the fact that the entropy of a continuous probability distribution is not bounded from below.}
\label{plt:entropy_formula_plot}
\end{center}
\end{figure}

When the symplectic eigenvalues $\{ d_i \}$ are large (which in a thermal state can be shown to correspond to a high temperature limit for thermal states), the expression $S(d)$ becomes
\begin{equation}
S(d) \approx 1 + \log(d),
\end{equation}
so the quantum entropy formula approaches the classical formula \eqref{eq:classical_entropy} obtained above.
This fact is illustrated in figure~\ref{plt:entropy_formula_plot}, which shows the classical and quantum formulas for $S(d)$.

\section{Von Neumann entropy}
\label{section:ee}

We now calculate the entropy associated to a uniformly spaced lattice of field samples.
There are three distinct regimes, depending on whether the spacing is equal to the Nyquist spacing, larger than Nyquist (undersampling), or smaller than Nyquist (oversampling).
We will consider each of these possibilities in turn, both in vacuum and at finite temperature.

\subsection{Nyquist sampling}
\label{subsection:nyquist}

As shown in section \ref{section:gaussian}, the entropy of a set of sample points in a Gaussian state is determined by the commutators and two-point functions at the sample points.
We now calculate these for a quantum field at finite temperature.

\begin{figure}[ht]
\begin{center}
\includegraphics[height=2.5in,width=3.3in]{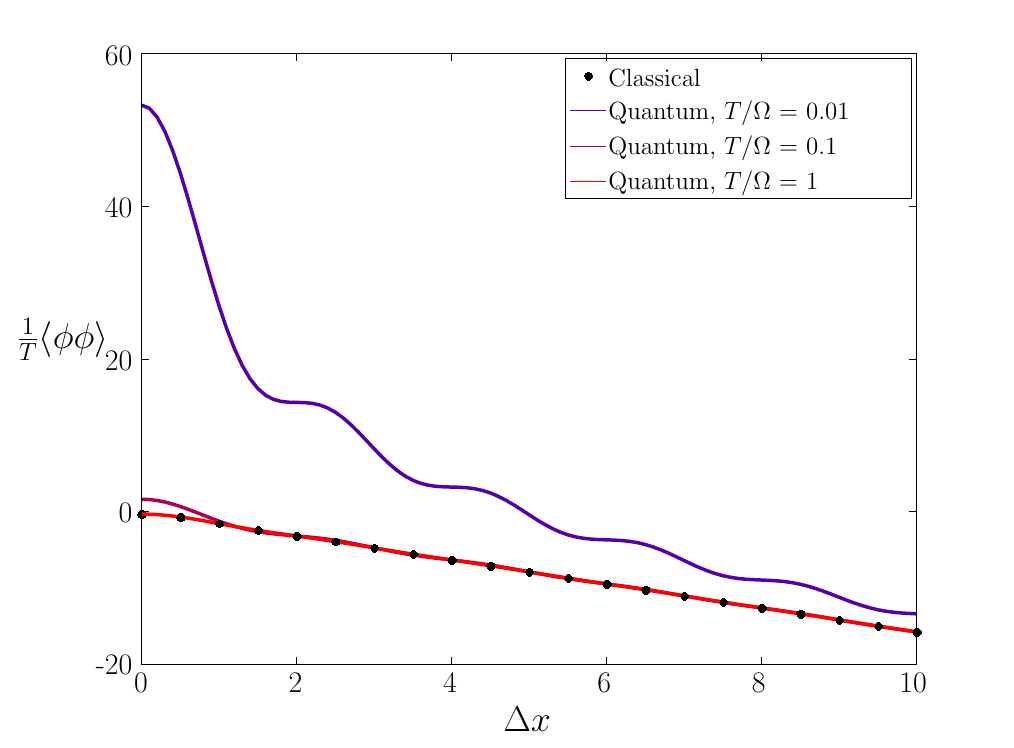}
\caption{\small
  Classical and quantum $\phi$-$\phi$ correlations as a function of their separation at various temperatures.
  The horizontal axis is scaled by $\Omega/\pi$ so that integer values correspond to Nyquist spacings.
  The vertical axis is scaled by $\Omega/T$.
  The classical correlator is plotted as black dots.
  The scaling of the vertical axis absorbs all of the temperature dependence of the classical correlator (since it grows proportionally to $T/\Omega$), thus the single graph of the classical correlator completely characterises its behaviour.
  The quantum correlators are shown as lines for temperatures up to $T/\Omega = 1$, at which point the quantum correlator converges to the classical correlator.
}
\label{plt:hightemp_phicorrlns}
\end{center}
\end{figure}

\begin{figure}[ht]
\begin{center}
\includegraphics[height=2.5in,width=3.3in]{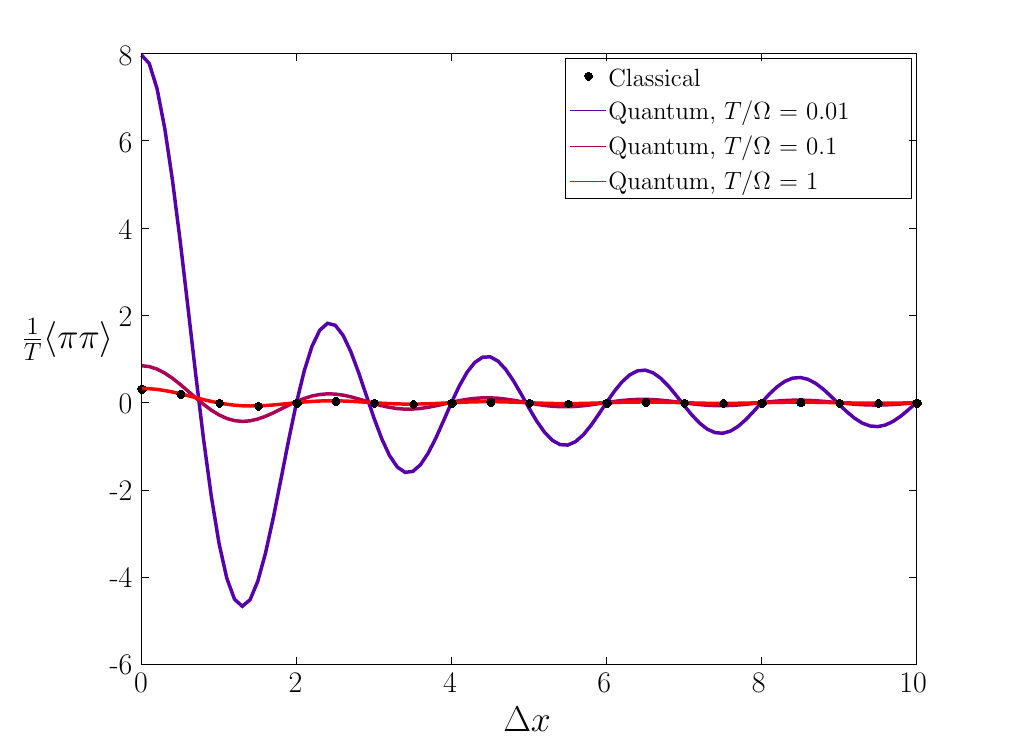}
\caption{\small
  Classical and quantum $\pi$-$\pi$ correlations as a function of their separation at various temperatures.
  The horizontal axis is scaled by $\Omega/\pi$ so that integer values correspond to Nyquist spacings.
  The vertical axis is scaled by $\Omega/T$.
  The classical correlator is plotted as black dots.
  The scaling of the vertical axis absorbs all of the temperature dependence of the classical correlator (since it grows proportionally to $T/\Omega$), thus the single graph of the classical correlator completely characterises its behaviour.
  The quantum correlators are shown as lines for temperatures approaching $T/\Omega = 1$, at which point the quantum correlator converges to the classical correlator.
}
\label{plt:hightemp_picorrlns}
\end{center}
\end{figure}

Before we perform the tracing operation, the density matrix associated with a single mode of the field is
\begin{equation}
\rho_k = \frac{1}{Z_k} \sum_{n_k = 0}^\infty e^{-\beta \omega_k \left( n_k + \frac{1}{2} \right)} \ket{n_k} \bra{n_k}.
\end{equation}
Formally, the total density matrix for the field is
\begin{equation}
\rho = \bigotimes_{\omega < \abs{k} < \Omega} \rho_k.
\end{equation}
This leads to the power spectra
\begin{eqnarray}
\langle | \phi_k |^2 \rangle = \mbox{tr} (| \phi_k |^2 \rho) &=& \frac{1}{\omega_k} \left( \frac{1}{e^{\beta \omega_k} - 1} + \frac{1}{2} \right), \\
\langle | \pi_k |^2 \rangle = \mbox{tr} (| \pi_k |^2 \rho) &=& \omega_k \left( \frac{1}{e^{\beta \omega_k} - 1} + \frac{1}{2} \right).
\end{eqnarray}
From the power spectra we find the two-point functions for a massless bandlimited field are
\begin{eqnarray}
\langle \phi(x) \phi(x') \rangle &=& \int_{\omega}^{\Omega} \frac{dk}{\pi} \cos(k\Delta x) \frac{1}{k} \left( \frac{1}{e^{\beta k} - 1} + \frac{1}{2} \right), \\
\langle \pi(x) \pi(x') \rangle &=& \int_{\omega}^{\Omega} \frac{dk}{\pi} \cos(k\Delta x) k \left( \frac{1}{e^{\beta k} - 1} + \frac{1}{2} \right).
\end{eqnarray}
These correlators are plotted in Figures~\ref{plt:hightemp_phicorrlns} and \ref{plt:hightemp_picorrlns} alonside the classical correlators.
We notice that, as expected, the classical and quantum correlators agree at high temperatures.
Both classical and quantum correlations exhibit oscillations at the ultraviolet scale.
At temperatures $T/\Omega \lesssim 1$, the quantum field exhibits stronger correlations than its classical counterpart due to vacuum fluctuations.

\begin{figure}[ht]
\begin{center}
\includegraphics[height=2.5in,width=3.3in]{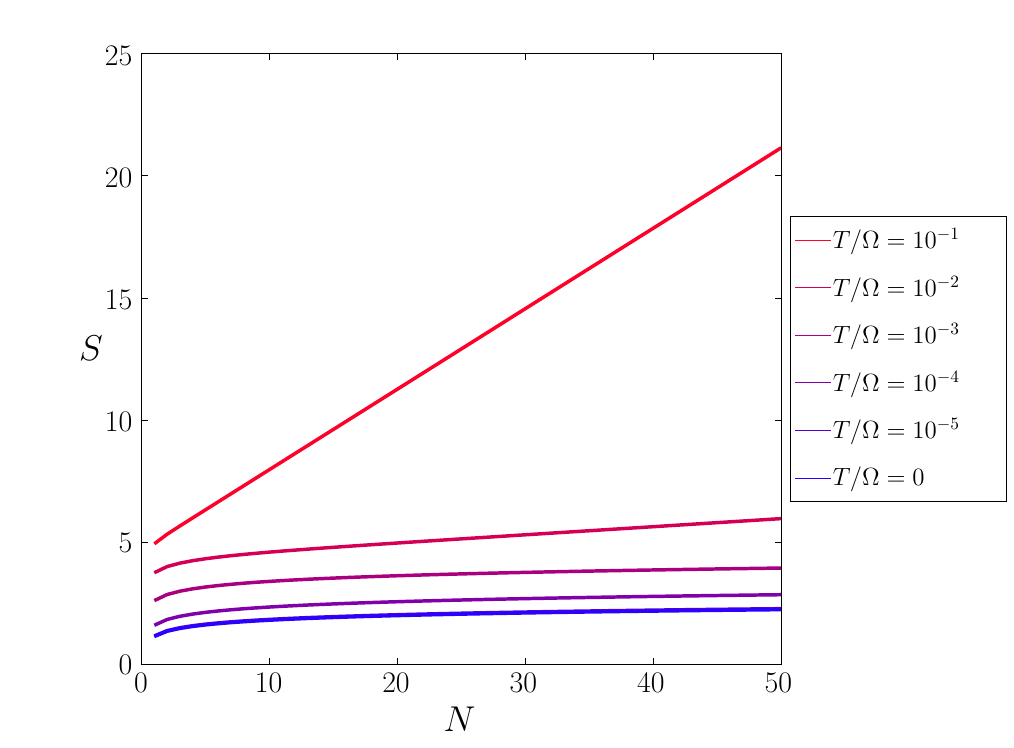}
\caption{\small
Entropy for $N$ Nyquist spaced sample points for a thermal state of a quantum Klein-Gordon field.
The plot illustrates the transition between the logarithmic behaviour at low temperature to linear behaviour at higher temperatures where the entropy becomes an extensive variable.}
\label{plt:Sq_thermal}
\end{center}
\end{figure}

Using these thermal correlation functions together with the commutation relations (eq.~\eqref{eq:commutator}), we numerically calculate the entanglement entropy of a finite lattice of points separated by the Nyquist spacing $\pi / \Omega$.
Then we consider the change in the entropy as the number of points increases.
The entropy as a function of the number of sample points is illustrated in Figure~\ref{plt:Sq_thermal} for a range of temperatures.
The plot shows clearly the transition between linear and logarithmic growth of the entropy.

Comparing the entropy in the quantum theory (Fig.~\ref{plt:Sq_thermal}) with the corresponding classical result (Fig.~\ref{plt:Scl_thermal}), 
we see the expected agreement of the scaling behaviour at high temperatures, with the entropy growing linearly with the number of lattice points.
This simply reflects the extensivity of the thermal entropy.
However, at low temperatures the entropy of the quantum field increases only logarithmically with the number of points removed (e.g. the zero temperature state in Figure~\ref{plt:Sq_thermal}).
This is a well-known result for conformal field theories, such as the massless scalar field, for which the entropy scales as $S \sim \frac{1}{3} \log (\Omega L)$ \cite{Holzhey1994}.
Here, for the thermal state at zero temperature (i.e., the ground state) the fitted curve to the data is
\begin{equation}
S(N) = 0.334 \log N + 3.25
\end{equation}
which is in agreement with the expected leading order term $\frac{1}{3}\log(N)$.
This result is for an infrared scale of $\omega/\Omega = 10^{-300}$.

We therefore see that the total entropy of the quantum field is a combination of the thermal and entanglement entropy.
At high temperatures the entropy is primarily thermal, while at low temperatures the thermal entropy vanishes, leaving only the entanglement entropy.

\subsection{Undersampling}
\label{subsection:undersampling}

Now we examine how the entanglement entropy of the ground state of the quantum field scales with the number of samples $N$ when the distance between adjacent samples is independent of $\Omega$.
In sampling theory terminology, this corresponds to undersampling when $\Delta x > \pi / \Omega$ and oversampling when $\Delta x < \pi / \Omega$.

First, we will examine the case of undersampling.
Recall that the strength of correlations between the field at two separate points decays with the distance between the points and the entropy is dominated by local correlations.
Thus, for points separated by more than the Nyquist spacing any given point is most strongly correlated with degrees of freedom in the complementary subsystem.
As a result, the entanglement entropy in this regime is proportional to the number of points traced out.
Hence we should recover a volume law for the entropy similar to the high temperature thermal state.

\begin{figure}[ht]
\begin{center}
\includegraphics[height=2.5in,width=3.3in]{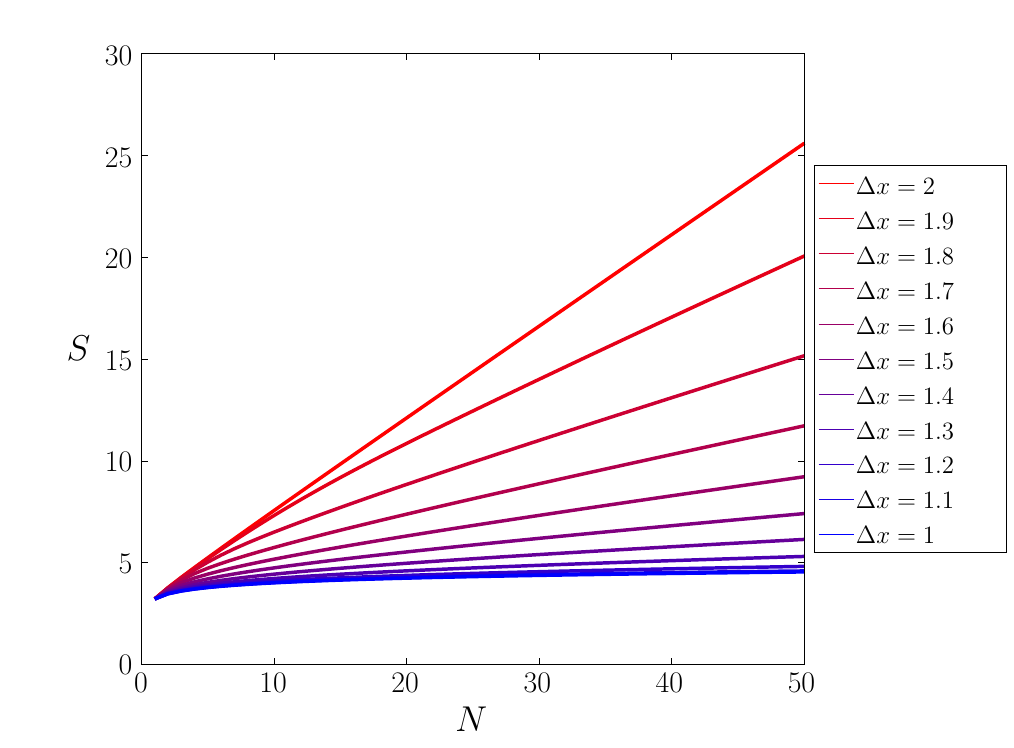}
\caption{\small
  The dependence of the entanglement entropy on the number of sample points, for sample spacings between 1 and 2 Nyquist spacings.
  The $\Delta x$ labels in the legend are scaled by $\Omega/\pi$ so that the Nyquist spacing is 1.
  We see that once the adjacent point spacing has reached twice the Nyquist spacing, the entropy has transitioned from the logarithmic scaling law to a volume law.
  The infrared scale is $\omega/\Omega = 10^{-300}$.
}
\label{plt:undersampling}
\end{center}
\end{figure}

The procedure to calculate the entanglement entropy is the same as before, except now the correlators and commutators are taken between points which do not all lie on a Nyquist lattice.
Figure~\ref{plt:undersampling} shows the dependence of the entanglement entropy on the number of points traced out for spacings that vary between 1 and 2 Nyquist spacings.

We see that for Nyquist and near-Nyquist sample spacings the entanglement entropy grows logarithmically with the number of points. As the sample separation increases there is a transition to a linear scaling with the number of points.
This corroborates our intuition for the scaling behaviour of the entanglement entropy when undersampling.

\subsection{Oversampling}
\label{subsection:oversampling}

Instead of separating the sample points, we now take $N$ samples which are spaced closer than a Nyquist spacing apart.
Starting from Nyquist spacing and pushing the points closer together, we find that the entropy does not depend very sensitively on the spacing between points.
The logarithmic scaling of the entanglement entropy remains the same, and only the subleading constant is modified.
This is in contrast with the case of undersampling, where we saw a transition from logarithmic to linear growth.

Unfortunately, with an oversampled set of lattice points the above numerical calculation becomes unstable for a large number of sample points.
This is because the matrices $\Lambda$ and $\Sigma$ defined in section \ref{subsection:quantum} become ill-conditioned.
However, it is possible to perform these numerical calculations in the regimes of a small amount of oversampling, or for only a small number of points.

For a small amount of oversampling, that is, with lattice spacing between 97\%-100\% of the Nyquist spacing, we continue to find logarithmic scaling of the entanglement entropy with the number of points traced out:
\begin{equation}
S(N) = c_0 \log N + c_1
\end{equation}
where in each case $c_0 \in [0.333, 0.335]$ and $c_1 = 3.25$.
This result is for an infrared scale of $\omega/\Omega = 10^{-300}$.

\begin{figure}[ht]
\begin{center}
\includegraphics[height=2.5in,width=3.3in]{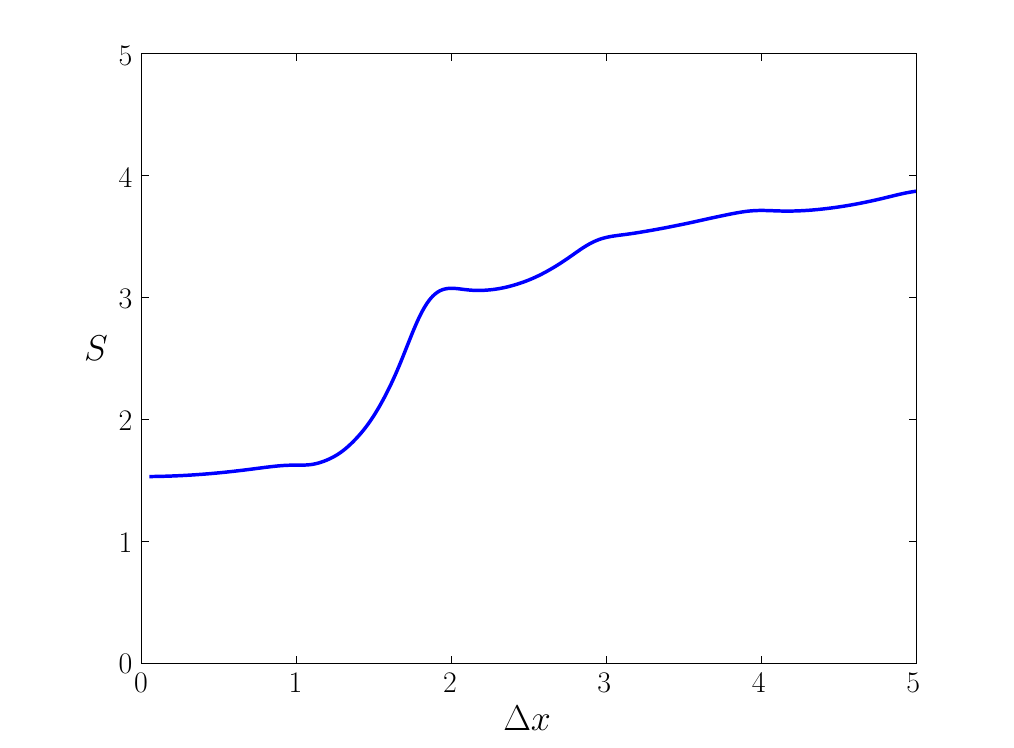}
\caption{\small
  Entanglement entropy of five points as a function of the spacing between them.
  The horizontal axis is scaled by $\Omega/\pi$ so that the Nyquist spacing is 1.
  For spacings below the Nyquist spacing, we see a plateauing effect indicating that the entanglement entropy is not sensitive to the spacing between the points for spacings below the Nyquist spacing.
  Above the Nyquist spacing the entropy tends to increase as the spacing increases.
  In this plot the infrared to ultraviolet ratio is $\omega/\Omega = 10^{-5}$.
}
\label{plt:ground_plateauing}
\end{center}
\end{figure}

Although we were not able to extend far into the oversampling regime, if we fix a small number of points it is possible to calculate the entanglement entropy of these points for an arbitrary amount of oversampling.
Figure~\ref{plt:ground_plateauing} shows the entanglement entropy for 5 sample points as a function of the separation between the points.
One sees from the figure that the entanglement entropy reaches a plateau for spacings below the Nyquist spacing, with the entropy depending only weakly on the spacing in this regime.
This shows that the logarithmic scaling law for the entanglement entropy is also independent of the spacing between the points, if the spacing is below the Nyquist spacing.

In figure~\ref{plt:ground_plateauing} we also see that the entropy continues to increase as the spacing is increased above the Nyquist spacing.
This is because the entropy depends on local correlations, and as the degrees of freedom become more separated, more of their correlations are with the complementary degrees of freedom.
In the 1+1 field theory we consider, these correlations decay very slowly, so that the entropy continues to increase all the way up to the infrared cutoff scale.
The entropy is also slightly peaked at integer multiples of the Nyquist spacing: this is related to the fact that the commutators vanish between degrees of freedom separated by multiples of the Nyquist spacing.

\begin{figure}[ht]
\begin{center}
\includegraphics[height=2.5in,width=3.3in]{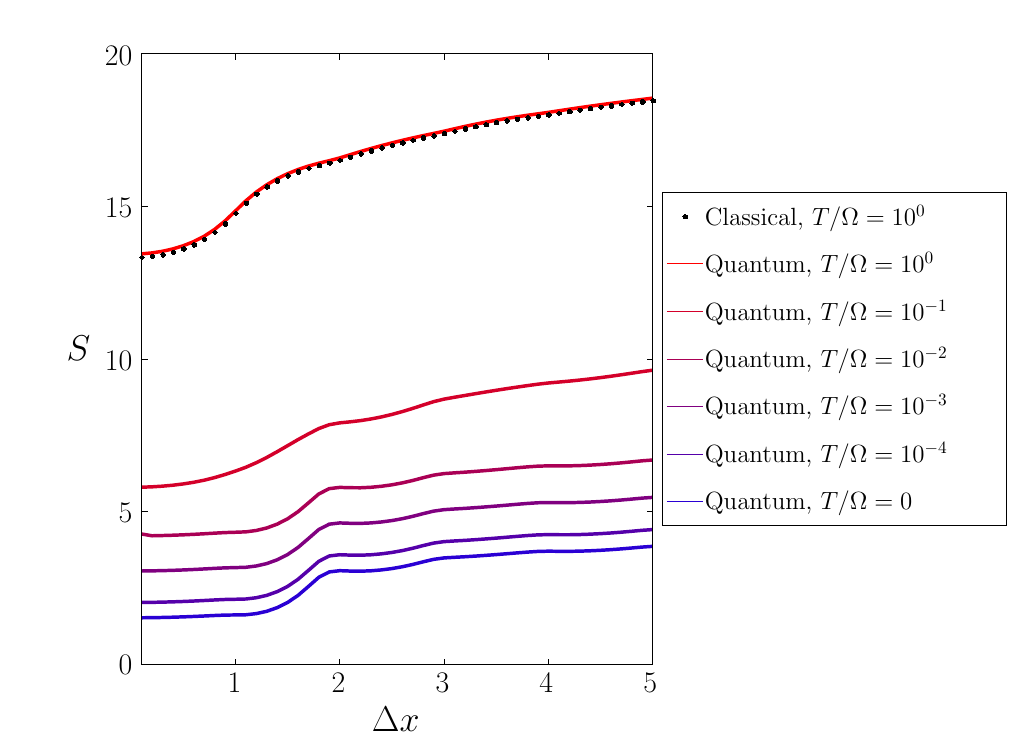}
\caption{\small
  Entanglement entropy of five points as a function of the spacing between them at various temperatures.
  The horizontal axis is scaled by $\Omega/\pi$ so that the Nyquist spacing is 1.
  We see here a plateau in the entropy at small spacings, similar to the plateau in the ground state entropy.
  The analogous classical calculation is also shown as black dots for temperature $T/\Omega = 1$.
  At temperatures $T/\Omega > 1$, the entropy behaves the same as for $T/\Omega = 1$ but shifted vertically by a constant $\sim \log(T/\Omega)$.
  In this plot the infrared to ultraviolet ratio is $\omega/\Omega = 10^{-5}$.
}
\label{plt:thermal_plateauing}
\end{center}
\end{figure}

We can also show that the plateau in the entropy at small spacings is not only a feature of the ground state, but also occurs in the thermal state considered in section \ref{subsection:nyquist}.
Figure~\ref{plt:thermal_plateauing} shows the entanglement entropy of five points as a function of their spacing for various temperatures.
We see that at any temperature, when the points are closer than a Nyquist spacing, the resulting entanglement entropy depends only weakly on the spacing.
This plateau is therefore not a fundamentally quantum-mechanical effect: if we perform the same calculation with five points for the classical thermal state, we find the same plateau.

The restriction to a small number of sample points is a consequence of numerical instability that occurs for small spacing and large numbers of points.
In the following subsection we develop an analytic method that will allow us to consider a larger number of points, in the limit of small spacing.

\subsection{Oversampling using derivatives}

The above results suggest that even if the $N$ sample points are arbitrarily close together, the corresponding entanglement entropy will scale as $\frac{1}{3} \log (N)$.
However, these results are restricted either to a small number of lattice points, or to very mild oversampling.
To show this scaling behaviour we will develop a method that allows us to consider a larger number of sample points in the limit of small spacing.
In the limit where the $N$ points are taken coincident, sampling at $N$ points is equivalent to sampling the first $N$ spatial derivatives at a single point.
This is related to the result that in classical sampling theory that instead of sampling the field at the Nyquist rate, one can instead sample the field and its derivatives at a fraction of the Nyquist rate \cite{Higgins1999}.
For example, if at a sample point one measures both the amplitude of the field and its first derivative, then the samples only need to be taken at half the Nyquist rate.

First, consider a situation where we sample at just two nearby points.
We can perform the following symplectic transformation on the phase space variables:
\begin{equation}
\left( \begin{array}{c} \phi(x) \\ \phi(x+\Delta x) \\ \pi(x) \\ \pi(x+\Delta x)\end{array} \right) \to \left( \begin{array}{c} \frac{1}{\sqrt{2}} ( \phi(x+\Delta x) + \phi(x) ) \\ \frac{1}{\sqrt{2}} ( \phi(x+\Delta x) - \phi(x) ) \\ \frac{1}{\sqrt{2}} ( \pi(x+\Delta x) + \pi(x) ) \\ \frac{1}{\sqrt{2}} ( \pi(x+\Delta x) - \pi(x) ) \end{array} \right).
\end{equation}
Thus, we see that sampling two points in the limit where their separation vanishes, $\Delta x \to 0$, can equivalently be viewed as sampling the field and its derivative at a single point (as well as the conjugate momentum and its derivative at that point).
This motivates performing the calculation of the entanglement entropy after sampling the field and its higher derivatives $\lbrace \phi, \phi', \phi'', \dots, \phi^{(N-1)} \rbrace$ at a single point, which is equivalent to sampling the field at $N$ points in the limit of small spacing.
This can be thought of as an extreme case of oversampling the field.

The required matrix elements for the calculation of the entropy are:
\begin{widetext}
\begin{equation}
\left[ (\partial_x)^n \phi (x), (\partial_x)^m \pi (x) \right] = \begin{cases} \frac{i}{\pi} (-1)^{\frac{n+3m}{2}} \frac{1}{n+m+1} \left( \Omega^{n+m+1} - \omega^{n+m+1} \right) &\mbox{ if } (n+m) = 0 \mod 2 \\
0 &\mbox{ if } (n+m) = 1 \mod 2 \end{cases},
\end{equation}

\begin{equation}
\langle (\partial_x)^n \phi (x) \cdot (\partial_x)^m \phi (x) \rangle =  \begin{cases} (-1)^{\frac{n+3m}{2}} \frac{1}{2\pi} \log \left( \frac{\Omega}{\omega} \right)  &\mbox{ if } n+m=0 \\
(-1)^{\frac{n+3m}{2}} \frac{1}{2\pi} \frac{1}{n+m} \left( \Omega^{n+m} - \omega^{n+m} \right) &\mbox{ if } n+m \neq 0, (n+m) = 0 \mod 2 \\
0 &\mbox{ if } (n+m) = 1 \mod 2 \end{cases},
\end{equation}
and 
\begin{equation}
\langle (\partial_x)^n \pi (x) \cdot (\partial_x)^m \pi (x) \rangle =  \begin{cases} (-1)^{\frac{n+3m}{2}} \frac{1}{2\pi} \frac{1}{n+m+2} \left( \Omega^{n+m+2} - \omega^{n+m+2} \right) &\mbox{ if } (n+m) = 0 \mod 2 \\
0 &\mbox{ if } (n+m) = 1 \mod 2 \end{cases}.
\end{equation}
\end{widetext}

\begin{figure}[ht]
\begin{center}
\includegraphics[height=2.5in,width=3.3in]{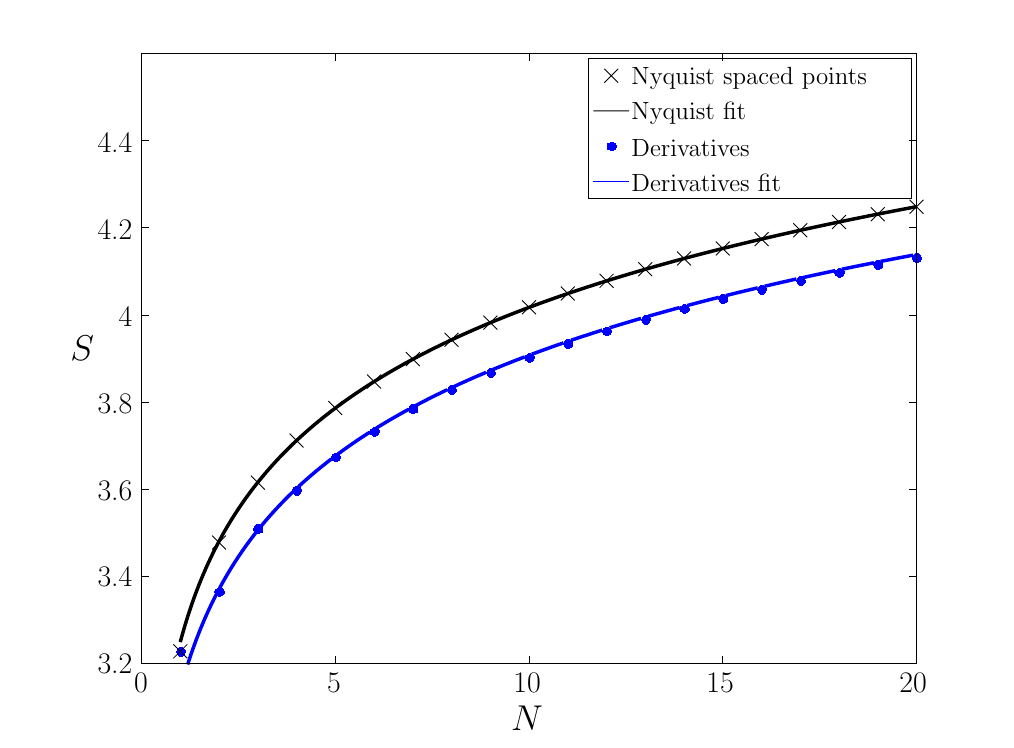}
\caption{\small
  Entanglement entropy dependence of number of derivatives traced out at a single point.
  Fitted curve to Nyquist spaced points is $S(N) = 0.334 \log(N) + 3.25$, where $N$ is the number of sampling points.
  Fitted curve to derivative sampling points is $S(N) = \tfrac13 \log(N) + 3.14$, where $N-1$ is the number of derivatives sampled (which corresponds to $N$ sampled points).
  We see that both curves differ by a constant $\approx 0.11$.
  The infrared to ultraviolet ratio is $\omega/\Omega = 10^{-300}$.
}
\label{plt:derivs}
\end{center}
\end{figure}

Figure \ref{plt:derivs} shows the entropy as a function of the number of points in the limit of small spacing.
For reference we also show the entropy for Nyquist spacing.
Both results are in agreement with the curve $1/3 \log(N) + c$, but with different constants.

This is consistent with the interpretation that both Nyquist spacing and the small spacing limit calculate the entropy of an interval of the same length, but regulated in a slightly different way.
This different regularization in the small spacing limit slightly disentangles the degrees of freedom very close to the entangling surface.
This reduces the constant coefficient in the entropy, while keeping the leading logarithmic behaviour.
We will give some evidence for this interpretation in section \ref{section:localization}.

\section{Localization of Degrees of Freedom}
\label{section:localization}

We have shown how the entanglement entropy of a set of samples of a bandlimited quantum field depends on the number of samples taken and on the spacing between samples.
For spacings larger than the Nyquist spacing the entanglement grows linearly with the number of samples.
At the Nyquist spacing there is a sharp transition.
For samples spaced at a Nyquist spacing or closer, the entanglement entropy grows logarithmically with the number of samples, with a coefficient that is independent of the spacing.
We will now show how this is a consequence of each mode of the field occupying an incompressible volume of space, which we will call a Planck volume. 

Since the degrees of freedom on all sample points of a Nyquist lattice is equivalent to the entire field, one can interpret each degree of freedom on the Nyquist lattice as representing information roughly contained in a Planck volume centered at the sample point.
Thus, we can interpret sampling a set of points on a Nyquist lattice as sampling the corresponding interval of the field.
If we sample finitely many contiguous degrees of freedom on a Nyquist lattice, the resulting entanglement entropy is generated by the correlations cut across the boundary of the interval.
Thus, the entanglement entropy varies logarithmically with the number of points sampled, $N$, since they describe an interval of length $L = N\pi/\Omega$.

If the sample points we take are spaced farther than a Nyquist spacing, the corresponding Planck volumes centered at these sample points will be disjoint.
Therefore, the local correlations around each point which contribute to the entanglement entropy will mostly be independent.
Therefore, the entanglement entropy for points which are well-separated will scale linearly with the number of sampled degrees of freedom, as we demonstrated above.

The most interesting case is the case of oversampling.
When the sampled degrees of freedom are pushed very close together, we find that the entropy varies logarithmically with the number of degrees of freedom sampled.
As the degrees of freedom are taken together one might expect to probe only a single Planck volume, and hence find a smaller entropy.
If the $N$ points are almost on top of one another, the correlations at the boundary of the overlapping Planck volumes would generate entanglement entropy roughly equivalent to the entanglement entropy of a single sampled degree of freedom, rather than $N$ degrees of freedom.
However, from the results of the entropy calculation, it seems that the $N$ degrees of freedom correspond to a larger, effective volume the size of the original volume described by the $N$ contiguous Nyquist-spaced points before they were pushed together.

We can estimate the size of this effective volume as follows.
First, we take a set of $N$ sample points with some equidistant spacing, which we shall denote subsystem $A_N$.
Now we include another sample point which we initially place very far from these $N$ sample points, which we shall denote subsystem $P$.
This point will be used as a probe to localize the points in subsystem $A_N$, using the mutual information
\begin{equation}
  I(A_N:P) = S(A_N) + S(P) - S(A_N,P),
\end{equation}
where $S(A_N)$, $S(P)$ are the entanglement entropies of the subsystems $A_N$ and $P$ (respectively), and $S(A_N,P)$ is the entanglement entropy of the combined system.
While the probe point $P$ remains far away, the mutual information will be small because the correlations between $A_N$ and $P$ are small.
When $P$ approaches the system of $N$ points, the mutual information will increase as the two subsystems become more correlated.
We can use the increase of the mutual information $I(A_N:P)$ as an indicator of the extent of the subsystem $A_N$.
In particular, we can map the boundary of $A_N$ for various values of the spacing between the $N$ points.
If the points spaced far below the Nyquist spacing indeed describe an effective volume of $\sim N$ Planck volumes in size, then we should see the boundary of the subsystem at the edge of this region.

\begin{figure*}[ht]
\begin{center}
\scalebox{0.7}{\includegraphics{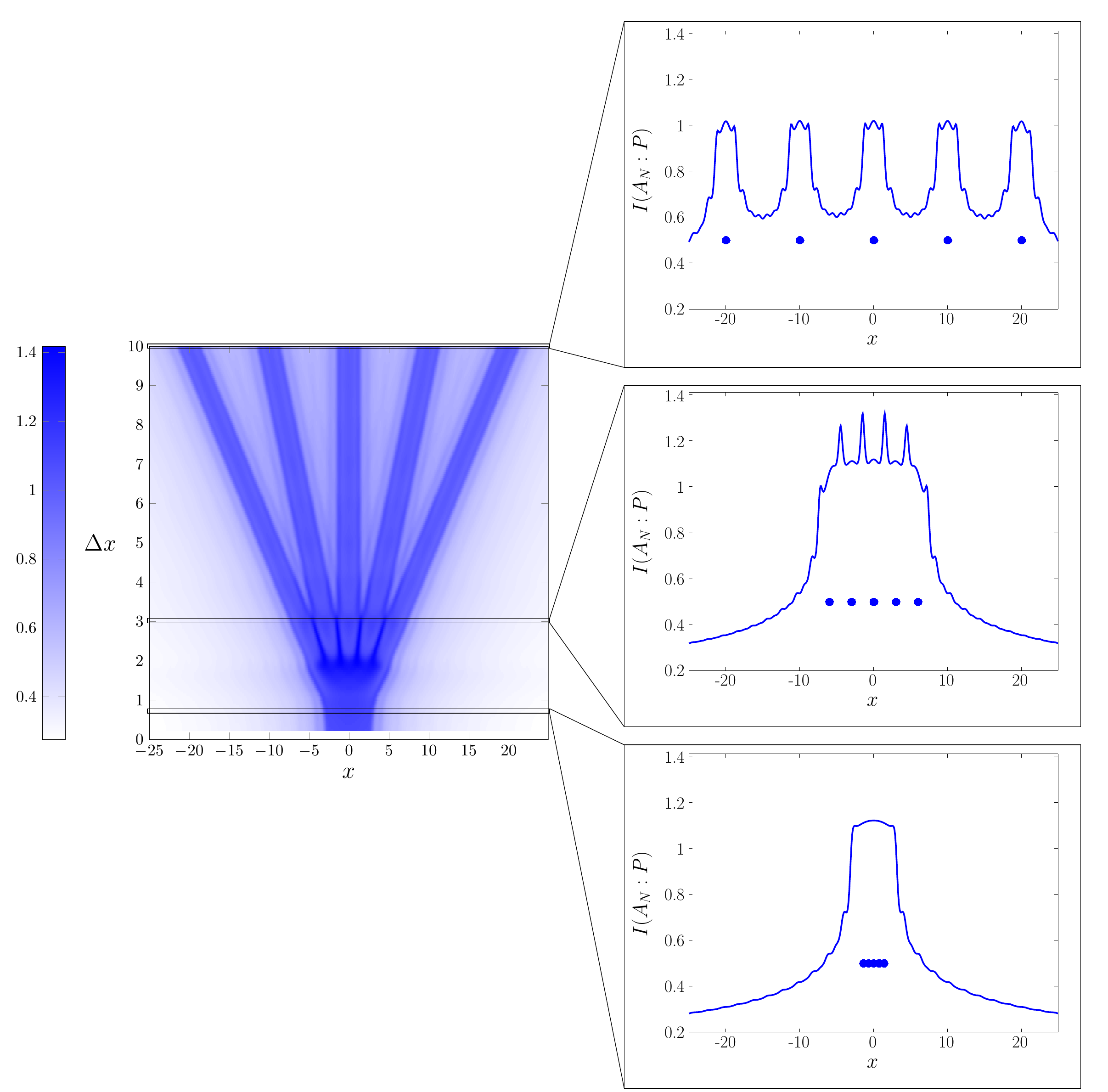}}
\caption{\small
  Mutual information between subsystem $A_N$ and probe point $P$.
  The spatial axes are scaled so that the Nyquist spacing is 1.
  The variable $x$ denotes the position of the probe point $P$, with $x=0$ at the centre of the subsystem $A_N$.
  For spacings much larger than the Nyquist spacing, the degrees of freedom occupy independent intervals of the order of the Nyquist spacing in size.
  For spacings at or below the Nyquist spacing, the points in the interval describe a fixed volume of size $N$.
}
\label{plt:explorer_mutinfo}
\end{center}
\end{figure*}

In figure~\ref{plt:explorer_mutinfo} we see this is the case for a subsystem of $N=5$ points.
When the points in the interval are spaced farther than a Nyquist spacing (in the upper region of the graph), we can clearly identify the region of space that they occupy, which is centered at each sample point and on the order of a Planck volume in size.
When the spacing approaches the Nyquist spacing, the points begin to occupy a single interval of roughly $N=5$ Planck volumes in size.
As the spacing decreases below the Nyquist spacing, the size of this interval ceases to decrease, indicating that the sample points describe the same volume of space regardless of their positions below the Nyquist spacing.
Therefore, the sample points each describe an independent volume of space of Planckian size.
In this sense, the Planck volumes described by the individual degrees of freedom are incompressible.

This behavior shows how the field is able to have a finite information density, while still allowing arbitrarily closely spaced probes.
If we attempt to sample the field at two points closer than a Nyquist spacing, we are really only probing the degrees of freedom in a larger region of space centered around these samples, whose volume is determined by the number of samples rather than by their spacing.

\section{Infrared behaviour}
\label{section:ir}

Above we have examined the effect of the ultraviolet cutoff on the calculation of the entanglement entropy of a bandlimited quantum field.
Here we also briefly examine the infrared behaviour of the entanglement entropy.
Refs. \cite{Casini2009,Mallayya2014} calculated the entanglement entropy of a free scalar field in 1+1 dimensions with an infrared cutoff.
They find that the entanglement entropy of a subset of $N$ oscillators takes the form
\begin{equation}
S = c_0 \log(N) + \frac{1}{2} \log \left( \log \left( \frac{1}{\omega} \right) \right) - \log \pi,
\end{equation}
where $c_0$ is a constant and $\omega$ is the infrared cutoff.
Here the entropy diverges as a double logarithm with the infrared cutoff.
This infrared divergence is a special feature of 1+1 dimensions and arises due to a build-up of long range correlations in the field.
We saw in section \ref{section:sampling} that the $\phi$-$\phi$ correlations decay very slowly with the point separation, and are only suppressed on the order of the longest wavelength $2\pi / \omega$.
If we remove the infrared cutoff, we allow for infinitely long wavelengths, and thus divergent correlations.

In our above numerical results for calculating the entanglement entropy of a quantum field on a Nyquist lattice, we perform a numerical fit of the data to determine the coefficients of the equation
\begin{equation}
S(N) = c_1 \log N + c_2.
\end{equation}
This numerical fit was also performed for various values of the infrared cutoff, $\omega / \Omega \in [10^{-50},10^{-300}]$.
We find that the leading order behaviour of $c_2$ in this range is
\begin{equation}
c_2 = \frac{1}{2} \log \left( \log \left( \frac{\Omega}{\omega} \right) \right).
\end{equation}
The coefficient $c_2$ agrees with the result of Ref. \cite{Mallayya2014} (up to a constant).
Furthermore, in the limit $\omega \ll \Omega$, the infrared cutoff $\omega$ has the same effect on the entropy as a small mass \cite{Casini2009}.

It is possible to anticipate this result analytically by considering tracing out a system of one oscillator.
From the expressions found above for the two-point functions, we find the diagonal terms are
\begin{eqnarray}
\left[ \phi(x), \pi(x) \right] &=& i \frac{\Omega - \omega}{\pi},\\
\langle \phi^2 (x) \rangle &=& \frac{1}{2\pi} \log \left( \frac{\Omega}{\omega} \right), \mbox{and}\\
\langle \pi^2 (x) \rangle &=& \frac{1}{4\pi} \left( \Omega^2 - \omega^2 \right).
\end{eqnarray}
The symplectic eigenvalues of the covariance matrix are
\begin{equation}
\lambda_{\pm} = \pm \frac{\pi}{\Omega - \omega} \sqrt{ \frac{1}{8 \pi^2} \left( \Omega^2 - \omega^2 \right) \log \left( \frac{\Omega}{\omega} \right) }.
\end{equation}
Thus, the entanglement entropy is
\begin{equation}
S = ( \lambda_+ + \frac{1}{2} ) \log ( \lambda_+ + \frac{1}{2} ) - ( \lambda_+ - \frac{1}{2} ) \log ( \lambda_+ - \frac{1}{2} ),
\end{equation}
which to leading order in the limit $\omega/\Omega \to 0$ is
\begin{equation}
S \sim \frac{1}{2} \log \left( \log \left( \frac{\Omega}{\omega} \right) \right).
\end{equation}
We see directly that the outer logarithm is due to the entanglement entropy formula, and the inner logarithm arises because the diagonal elements of the $\phi$-$\phi$ correlation matrix diverge logarithmically in the infrared as
\begin{equation}
\langle \phi^2(x) \rangle = \int_\omega^\Omega \frac{dk}{2\pi} \frac{1}{k} \nonumber = \frac{1}{2\pi} \log \left( \frac{\Omega}{\omega} \right).
\end{equation}

One may wonder how the infrared behaviour of the entanglement entropy changes with a different implementation of the infrared cutoff.
In particular, we will examine how the entropy behaves when the field modes are discrete.
Consider a scalar field on an interval $[0,L]$ with periodic boundary conditions (up to a phase),
\begin{equation}
\phi(0) = e^{i\alpha} \phi(L),
\end{equation}
where $\alpha \in [0, 2\pi)$.
Expanding the field in a spatial Fourier series,
\begin{equation}
\phi(x) = \sum_{k} \frac{1}{\sqrt{\omega_k L}} e^{ikx} \phi_k
\end{equation}
where $\omega_k := |k|$, one obtains a discrete set of momentum modes  $\lbrace k_n = \frac{2\pi n - \alpha}{L} \rbrace_{n \in \mathds{Z}}$.
The ultraviolet cutoff will be imposed by restricting the set of allowable $n$ to $\{ n \in \mathds{Z} : \omega_{k_n} = |k_n| \leq \Omega \}$.
We see that for $\alpha > 0$ there will be no zero mode, so the $\phi$-$\phi$ correlations are finite, but there will be an infrared divergence as $\alpha \to 0$.
It is straightforward to find that the Hamiltonian for this Klein-Gordon field can be expressed as
\begin{equation}
H = \frac{1}{2} \sum_{n} \omega_{k_n} \left( \pi^2_{k_n} + \phi^2_{k_n} \right).
\end{equation}
Expanding in terms of the usual creation and annihilation operators,
\begin{align}
a_k := \frac{1}{\sqrt{2}} \left( \phi_k + i \pi_k \right),
\end{align}
one finds
\begin{eqnarray}
\left[ \phi(x), \pi(x') \right] &=& \frac{i}{L} \sum_{n} e^{ik_n (x-x')},\\
\langle \phi(x) \phi(x') \rangle &=& \frac{1}{2L} \sum_{n} \frac{1}{\omega_{k_n}} e^{ik_n (x-x')}, \mbox{and}\\
\langle \pi(x) \pi(x') \rangle &=& \frac{1}{2L} \sum_{n} \omega_{k_n} e^{ik_n (x-x')}.
\end{eqnarray}
These matrix elements are easy to calculate numerically since the ultraviolet cutoff imposes a finite number of $n$ to sum over.

Similar to the calculation performed above, we shall calculate the entanglement entropy associated with tracing out the field oscillator at a single point.
The relevant diagonal elements of the correlation matrices are:
\begin{eqnarray}
\left[ \phi(x), \pi(x) \right] &=& i \frac{2K}{L},\\
\langle \phi^2(x) \rangle &=& \frac{1}{2\alpha} + \frac{1}{4\pi} \left[ \psi \left(K-\frac{\alpha}{2\pi}+1 \right) \right. \nonumber \\
&&- \psi \left(1-\frac{\alpha}{2\pi} \right) + \psi\left (K+\frac{\alpha}{2\pi} \right) \nonumber \\
&&- \left. \psi\left( 1+\frac{\alpha}{2\pi} \right) \right], \\
\langle \pi^2(x) \rangle &=& \frac{\pi K^2}{L^2},
\end{eqnarray}
where $K := \Omega L / 2 \pi$, and $\psi(x)$ is the digamma function.
We see that now instead of a logarithmic infrared divergence of $\langle \phi^2(x) \rangle$, with a discrete set of modes the correlation function diverges as $1/\alpha$ as $\alpha \to 0$.
In this limit, the leading order behaviour of the symplectic eigenvalues is
\begin{equation}
\lambda_\pm \sim \frac{1}{2} \sqrt{\frac{\pi}{2\alpha}}.
\end{equation}
The corresponding entanglement entropy is
\begin{equation}
\label{eq:discreteir}
S \sim \frac{1}{2} \log \left( \frac{\pi}{2\alpha} \right) - \log 2 + 1.
\end{equation}
Numerical calculations were performed with multiple points traced out, and the observed behaviour in the limit $\alpha \to 0$ agrees with that in equation \eqref{eq:discreteir}.

For our ultraviolet and infrared bandlimited quantum field on the real line, the maximum allowable wavelength is $\lambda_{max} = 2\pi / \omega$.
For periodic boundary conditions, the maximum wavelength is $\lambda_{max} = L/2\pi \alpha$.
Thus, we can compare and summarise the infrared behaviour of the entanglement entropy as
\begin{itemize}
\item For a continuous set of modes:
\begin{align}
&S \sim \frac{1}{2} \log \left( \log \left( \Omega \lambda_{max} \right) \right) && \mbox{ as } \lambda_{max} \to \infty.
\end{align}
\item For a discrete set of modes:
\begin{align}
&S \sim \frac{1}{2} \log \left( \frac{\lambda_{max}}{L} \right) && \mbox{ as } \lambda_{max} \to \infty.
\end{align}
\end{itemize}
In the case of boundary conditions that are periodic up to a phase, the entropy diverges more quickly.
This is plausible since this case is a description of a field on a circle, and any two points on this circle coupled to one another an arbitrary number of times in the limit $\lambda_{max} \to \infty$.

\section{Outlook}
\label{section:summary}
Shannon sampling, which in information theory establishes the equivalence of continuous and discrete information, also yields a simple model for how a natural ultraviolet cutoff at the Planck scale might manifest itself in QFT. Namely, when approaching the Planck scale from lower energies, where QFT is still valid, quantum fields are modeled as being bandlimited. 

Within this model, we probed the localization of the quantum field's degrees of freedom by tracking the behaviour of the vacuum entanglement entropy.
We found that the degrees of freedom of the quantum field, i.e., the local field oscillators, are non-local but can be localized down to the Planck-scale. 
In fact, we found that the local degrees of freedom occupy incompressible Planck-scale volumes, in the sense that $N$ degrees of freedom always describe $N$ Planck volumes regardless of the distance between them.

The tools that we outlined in this paper are applicable also to the study of bandlimited quantum fields in higher dimensions, and to fermionic fields.

For the case of higher dimensions, we conjecture that we will continue to see incompressibility of the field degrees of freedom, i.e., a plateau in the entanglement entropy for spacings below the Nyquist spacing. A system of $N$ lattice points should continue to probe a region of $N$ Planck volumes even if they are sub-Nyquist spaced.
In higher dimensions it will be interesting to see how the shape of this region changes with the relative location of the sample points.
Moreover, at larger than Nyquist spacings, we expect to find an infrared plateau. This is because the vacuum correlations decay over shorter distances in higher dimensions and do, therefore, not exhibit the infrared divergences that soften the area scaling to a logarithmic scaling in 1+1 dimensions.

Our approach to studying localization and vacuum entanglement should also work for quantum fields on curved spacetimes. To this end, the generalization of Shannon sampling theory to curved spaces developed in \cite{Kempf2008} can be used. 
It will be interesting, in particular, to determine how the localizability of the fundamental field oscillator degrees of freedom is affected by curvature and horizons, and in particular, how this may affect Hawking radiation. 
A similar hard momentum cutoff was considered in \cite{Jacobson:2007jx}, and it should be interesting to apply our new methods to that case.

A further direction of great interest would be to study the entanglement entropy and the localization of degrees of freedom of fields with a fully covariant ultraviolet cutoff.
In this case, the bandlimitation would be imposed by cutting off the spectrum of a covariant Laplacian (on Riemannian manifolds) or a d'Alembert operator (on Lorentzian manifolds) \cite{Kempf2012}.
To this end, it may be useful to express the entropy purely in terms of spacetime correlation functions \cite{Sorkin2012}, i.e., without reference to the canonical formalism.
We remark that it has been argued that if the entanglement entropy is finite and obeys a form of the Clausius relation, then Einstein's equation emerges as a thermodynamic equation of state \cite{Jacobson1995,Jacobson2015}.
Thus having a regulator that is both covariant and cuts off the entanglement entropy could be a key step toward quantum gravity.

It should also be very interesting to study the interaction of Unruh-DeWitt detectors with the bandlimited quantum fields. Unruh deWitt detectors would describe explicit means to sample the quantum fields, and this could, therefore, constitute a further step towards the quantization of sampling theory. An interesting question that then arises is how the spatial profile of an Unruh-DeWitt detector interacts with the spatial profile of the degrees of freedom. Another key question that arises with the quantization of sampling theory follows from the fact that, in classical sampling theory, the reconstruction of a function from a lattice with non-equidistant sample points is more sensitive to noise in the sample measurements than reconstruction from a lattice with equidistant sample points. This, therefore, raises the interesting question if or to what extent quantum noise may interfere with the ability to use significantly irregularly-spaced lattices.

One of our central results is that the entanglement entropy does not decrease significantly as the spacing is decreased to below the Nyquist spacing.
In classical sampling theory too a curious phenomenon is known to arise as the spacing is decreased to below the Nyquist spacing, namely the phenomenon of superoscillations. 

Superoscillations in a bandlimited function are a finite set of oscillations that oscillate faster than the highest Fourier component in the signal \cite{Aharonov1990, Berry1994, Kempf:1999tq, Ferreira2002, Zheludev2008}.
Superoscillations are difficult to generate and rarely occur naturally but they have become an active field of investigation, also in engineering, because of their potential for superresolution. An open problem in this field is to quantify the prevalence of superoscillations in random signals. In our analogous classical calculation, we obtained probability distributions for observing particular field values on an arbitrary sampling lattice.
These distributions could be used to calculate the probability for finding the field in any particular superoscillatory configuration.
It might be possible to find in this way a practical measure of classical or quantum field fluctuations that is sensitive specifically to the occurrence of superoscillatory behaviour.

A key question that we addressed in this paper has been the question of how a bounded \it continuous \rm region can support only a finite number of degrees of freedom. In our model for a natural ultraviolet cutoff, this is accomplished by the quantum field degrees of freedom at distinct spacetime points not being independent. As a consequence, probing with too many operators in a small region simply causes one to probe a larger region of space. This means that the degrees of freedom of the theory are encoded redundantly on a single spatial slice. 

In fact it has recently been argued that, similarly, significant redundancy is present in the bulk field theory in AdS/CFT \cite{Almheiri2014}.
The redundancy implied by holography appears to be more extreme, however, since the number of degrees of freedom is bounded by area rather than volume.
To obtain this more drastic reduction, the nontrivial commutation relations coming from gravitation interaction may provide some clues \cite{Donnelly:2015hta}.
One may also take a cue from holography by trying to redundantly encode the degrees of freedom of a quantum field in a space of higher dimension. 

Finally, we note that the hyperbolic geometry of anti-de Sitter space suggests that an encoding based on wavelets may have some connection to holography \cite{Altaisky:2013xta,Qi:2013caa,White2013,Brennen2014}. Wavelet theory and Shannon sampling theory are closely related and the quantization of wavelet theory should be possible along the lines that we developed here. 

\section*{Acknowledgements}
AK and JP acknowledge support through the Discovery and NSERC-CGS-M (Canada Graduate Scholarship) programs of the 
Natural Sciences and Engineering Research Council (NSERC) of Canada.
JP also acknowledges support through the Ontario Graduate Scholarhip (OGS) program.

\vspace{1cm}

\appendix
\section{Functional analytic structure of sampling theory}
\label{app:fa_sampling}

Here we will briefly review the functional analytic structure of sampling theory, as first shown in  \cite{Kempf:2000fc}. 
The functional analytic view is powerful because it ultimately reduces sampling theory to the simple fact that when a Hilbert space vector is known in one basis then it is known in all bases. A key point here is the distinction between symmetric and self-adjoint operators.
By the spectral theorem, each self-adjoint operator possesses a unique diagonalization, whereas a simple symmetric operator does not, see, e.g., \cite{Akhiezer1993}. A certain type of simple symmetric operator, however, possesses a family of self-adjoint extensions, and their diagonalizations and spectra provide the sampling lattices of sampling theory. (The functional analytic view therefore also allows one  to work with varying Nyquist rates, which can occur in signal processing as well as in curved spacetimes.) 

Indeed, when acting upon the space of functions bandlimited by some maximum frequency $\Omega$, the usual position operator from first quantization, $\hat{X}$, is symmetric but not self-adjoint.
To see this, recall that for an operator $\hat{X}$ to be symmetric, it must obey
\begin{equation}
( \hat{X} \phi | \psi ) = ( \phi | \hat{X} \psi )
\end{equation}
for all $| \phi ), | \psi ) \in D(\hat{X})$, where $D(\hat{X})$ is the domain of $\hat{X}$. Equivalently, an operator is symmetric if and only if its expectation value is real for all vectors in its domain: $(\phi \vert X \phi)\in {\cal R} ~~\forall \phi \in D(\hat{X})$.  
For an operator to be self-adjoint, it must be symmetric and its domain must coincide with the domain of its adjoint, i.e., $( \hat{X} \phi | \psi ) = ( \phi | \hat{X} \psi )$ $\forall |\phi),|\psi) \in D(\hat{X})$ and $D(\hat{X}) = D(\hat{X}^\dagger)$.

Indeed, the position operator $\hat{X}$ with domain restricted to the space of bandlimited functions is symmetric, but the domain of its adjoint is a larger space of functions.
This is most easily seen in the momentum eigenbasis.
In this basis, the space of physical wavefunctions, and therefore the domain of $\hat{X}$, is the space of functions 
\begin{eqnarray}
D(\hat{X}) &=& \left\{ \phi\in L^2[-\Omega,\Omega]\vert \phi\in AC[-\Omega,\Omega],\right. \nonumber\\
& & \left. \phi'\in L^2[-\Omega,\Omega], \phi(-\Omega)=\phi(\Omega) = 0\right   \}
\end{eqnarray}
where $AC[-\Omega,\Omega]$ denotes the space of absolutely continuous functions.  
Notice that $D(\hat{X})$ contains only functions which obey Dirichlet boundary conditions.
In contrast, the domain of $\hat{X}^\dagger$ is the larger function space obtained by not imposing any boundary conditions.
This can be seen from the definition of $\hat{X}^\dagger$:
\begin{eqnarray}
\begin{split}
( \hat{X}^\dagger \phi | \psi ) =& ( \phi | \hat{X} \psi ) \\
=& \int dk \hspace{1mm} \tilde{\phi}^\ast (k) \left( i \frac{d}{dk} \tilde{\psi}(k) \right) \\
=& i \left( \tilde{\phi}^\ast (\Omega) \tilde{\psi} (\Omega) - \tilde{\phi}^\ast (-\Omega) \tilde{\psi} (-\Omega) \right) \\
&+ \int dk \hspace{1mm} \left( i \frac{d}{dk} \tilde{\phi} \right)^\ast (k) \tilde{\psi}(k).
\end{split}
\end{eqnarray}
We see from the last line that as long as $|\psi)$ is in the domain of $\hat{X}$ (thus obeys Dirichlet boundary conditions on the interval $[-\Omega,\Omega]$ in Fourier space), then we can define the adjoint of $\hat{X}^\dagger$ as $i(d/dk)$ in the momentum representation.
However, since we did not need to impose boundary conditions on $|\phi)$, we have that $D(\hat{X}) \subsetneq D(\hat{X}^\dagger)$.

Crucially now, it is possible to extend the domain of $\hat{X}$ so that the extension is a self-adjoint operator.
This operator is called a self-adjoint extension of $\hat{X}$ (e.g., see \cite{Fulling1989,Akhiezer1993}).
In our situation, we can perform this extension by enlarging the domain of $\hat{X}$ to include functions with periodic boundary conditions up to a phase on the interval $[-\Omega,\Omega]$ in Fourier space, i.e., $\tilde{\psi}(\Omega) = e^{-i\alpha} \tilde{\psi}(-\Omega)$.
We shall denote the self-adjoint extension corresponding to the particular phase $e^{-i\alpha}$ as $\hat{X}^{(\alpha)}$, thus the family of self-adjoint extensions is parametrized by $\alpha \in [0,2\pi)$.
Now we will show that the operators $\hat{X}^{(\alpha)}$ are indeed self-adjoint.
The definition of the adjoint of $\hat{X}^{(\alpha) \dagger}$ gives
\begin{eqnarray}
\begin{split}
( \hat{X}^{(\alpha) \dagger} \phi | \psi ) = &i \left( \tilde{\phi}^\ast (\Omega) \tilde{\psi} (\Omega) - \tilde{\phi}^\ast (-\Omega) \tilde{\psi} (-\Omega) \right) \\
&+ \int dk \hspace{1mm} \left( i \frac{d}{dk} \tilde{\phi} \right)^\ast (k) \tilde{\psi}(k) \\
= & i \left( \tilde{\phi}^\ast (\Omega)  - e^{i\alpha} \tilde{\phi}^\ast (-\Omega) \right) \tilde{\psi} (\Omega) \\
&+ \int dk \hspace{1mm} \left( i \frac{d}{dk} \tilde{\phi} \right)^\ast (k) \tilde{\psi}(k).
\end{split}
\end{eqnarray}
We see that the adjoint of $\hat{X}^{(\alpha)}$ is defined as $i(d/dk)$ in the momentum basis provided that $|\phi)$ also obeys the boundary condition $\tilde{\phi}(\Omega) = e^{-i\alpha} \tilde{\phi}(-\Omega)$.
Therefore we see $D( \hat{X}^{(\alpha)} ) = D( \hat{X}^{(\alpha) \dagger} )$, and so $\hat{X}^{(\alpha)}$ is self-adjoint.

Now, since the operators $\hat{X}^{(\alpha)}$ are self-adjoint, they have spectral decompositions.
The spectrum of the operator $\hat{X}^{(\alpha)}$ (for fixed $\alpha$) is discrete, $\spec (\hat{X}^{(\alpha)}) = \{ x_n^{(\alpha)} := \frac{2\pi n + \alpha}{2 \Omega} \}_{n \in \mathds{Z}}$, and describes a one-dimensional lattice.
The corresponding eigenvectors, $\{ |x_n^{(\alpha)}) \}_{n \in \mathds{Z}}$, are represented in the momentum eigenbasis as
\begin{equation}
( k | x_n^{(\alpha)} ) = \frac{1}{\sqrt{2\Omega}} e^{-i k x_n^{(\alpha)}}.
\end{equation}
These eigenvectors are orthogonal and admit a resolution of the identity:
\begin{equation}
\sum_{n \in \mathds{Z}} |x_n^{(\alpha)}) (x_n^{(\alpha)}| = \mathds{1}
\end{equation}.
Crucially, the position eigenvectors from different self-adjoint extensions are not orthogonal:
\begin{equation}
( x_n^{(\alpha)} | x_{n'}^{(\alpha')} ) = \sinc \left[ (x_n^{(\alpha)} - x_{n'}^{(\alpha')}) \Omega \right].
\end{equation}
Note that the union of the spectra of the entire family of self-adjoint extensions provides a covering of $\mathds{R}$.
Therefore, it is possible to construct an overcomplete continuum basis by taking the union of eigenbases of the family of self-adjoint extensions, i.e., $|x) := |x_n^{(\alpha)}) \iff x = x_n^{(\alpha)} := \frac{2\pi n + \alpha}{2 \Omega}$.

It is then simple to write down the Shannon sampling theorem for a bandlimited function $\psi$:
\begin{eqnarray}
\psi(x) &=&  ( x | \psi ) \\
&=& \sum_{n \in \mathds{Z}} ( x | x_n^{(\alpha)} )( x_n^{(\alpha)} | \psi ) \\
&=& \sum_{n \in \mathds{Z}} \sinc \left[ (x-x_n^{(\alpha)}) \Omega \right] \psi (x_n^{(\alpha)}).
\end{eqnarray}
Therefore, we see that the function $\psi$ is determined at any point $x \in \mathds{R}$ from its values on one of the lattices $\{ x_n^{(\alpha)} \}_n$.

Also, we obtain an overcomplete resolution of identity,
\begin{equation}
\frac{\Omega}{\pi} \int_\mathds{R} dx \hspace{1mm} |x)(x| = \mathds{1}
\end{equation}
where $\pi / \Omega$ is the density of degrees of freedom.
We can use the resolution of identity for the continuum basis to show that the space of bandlimited functions has a reproducing kernel,
\begin{eqnarray}
( x | \psi ) &=& \frac{\Omega}{\pi} \int dx' \hspace{1mm} ( x | x')( x' | \psi ) \\
\psi(x) &=& \int dx \hspace{1mm} K(x,x') \psi(x'),
\end{eqnarray}
where $K(x,x') := (\Omega/\pi) ( x | x' ) = (\Omega/\pi) \sinc [ (x-x') \Omega ]$.


\providecommand{\href}[2]{#2}\begingroup\raggedright\endgroup

\end{document}